\documentclass[aps,english,superscriptaddress]{revtex4-2}
\usepackage[utf8]{inputenc}
\usepackage{graphicx} 
\usepackage{amsmath}
\usepackage{epstopdf}
\usepackage{subfigure}
\usepackage{float}
\usepackage{slashed}
\usepackage{braket}
\usepackage{comment}
\usepackage{amssymb}
\usepackage{xcolor}
\usepackage{verbatim}

\begin{document}
\makeatletter
\newcommand{\rmnum}[1]{\romannumeral #1}
\newcommand{\Rmnum}[1]{\expandafter\@slowromancap\romannumeral #1@}
\makeatother
\title{Production of $K^+K^-$ Pairs Through Decay of $\phi$ Mesons}
\author{Xin-Nan Zhu}
\email{zxn19980402@mails.ccnu.edu.cn}
\affiliation{Institute of Particle Physics and Key Laboratory of Quark and Lepton Physics (MOS), \\
Central China Normal University, Wuhan 430079, China}
\affiliation{Institut f$\ddot{u}$r Theoretische Physik, Johann Wolfgang Goethe–Universit$\ddot{a}$t,
 Max-von-Laue-Str. 1, D–60438 Frankfurt am Main, Germany}
 \affiliation{Helmholtz Research Academy Hesse for FAIR, Campus Riedberg,
 Max-von-Laue-Str. 12, D–60438 Frankfurt am Main, Germany}

\author{Xin-Li Sheng}
\email{sheng@fi.infn.it}
\affiliation{Università degli studi di Firenze and INFN Sezione di Firenze,\\
Via G. Sansone 1, I-50019 Sesto Fiorentino (Florence), Italy}

\author{Defu Hou}
\email{houdf@ccnu.edu.cn}
\affiliation{Institute of Particle Physics and Key Laboratory of Quark and Lepton Physics (MOS), \\
Central China Normal University, Wuhan 430079, China}

\begin{abstract}
We develop a theoretical framework for the production of $K^+K^-$ pairs through the decay of $\phi$ mesons produced from a thermal background, based on the Nambu-Jona-Lasinio (NJL) model. The differential production rate of $K^+K^-$ is related to the self-energy of the $\phi$ meson, incorporating the contributions of the quark loop at leading order and the kaon loop at next-to-leading order in the $1/N_c$ expansion. We numerically evaluate the invariant mass spectrum of the $K^+K^-$ pair both in vacuum and at finite temperature. The inclusion of the kaon loop results in a finite width of the spectrum, improving agreement with experimental data. We also investigate the spin alignment of the $\phi$ meson induced by its motion relative to the thermal background. In the NJL model with only chiral condensates, we find that deviations from the unpolarized limit of 1/3 are negligible.
\end{abstract}

\maketitle

\section{Introduction}
In non-central heavy-ion collisions, the two initial nuclei carry a large amount of angular momentum, which is partially converted to the vorticity field in the quark-gluon plasma (QGP). The quarks in the QGP will be polarized through their spin-orbit coupling to the vorticity field and then combined into baryons or vector mesons with non-trivial polarizations \cite{Liang:2004ph, Liang:2004xn} (one can read, e.g. Ref. \cite{Florkowski:2018fap,Gao:2020lxh,Becattini:2020ngo,Becattini:2024uha,Niida:2024ntm,Chen:2024afy} for recent reviews). Therefore heavy-ion collisions provide a unique platform for phenomena related to spin. One typical example is the global spin polarization of $\Lambda$ hyperons measured by the STAR Collaboration at the Brookhaven National Laboratory (BNL) \cite{STAR:2017ckg,STAR:2018gyt}, which is believed to be evidence of the vorticity field and the large orbital angular momentum in the QGP. 

The vector meson's spin alignment, in some early works, is proposed to be induced by vorticity fields in a similar way as the $\Lambda$ polarization \cite{Liang:2004xn,Becattini:2016gvu,Yang:2017sdk}. In 2023, the STAR Collaboration observed a significant positive deviation from $1/3$ for the spin alignment of the $\phi$ meson \cite{STAR:2022fan}. Vorticity and magnetic fields can contribute to the spin alignment, as discussed in Refs. \cite{Sheng:2019kmk,Xia:2020tyd,Sheng:2022ssp,Wei:2023pdf,Kumar:2023ojl,DeMoura:2023jzz,Sun:2024anu,Chen:2024utf,Zhang:2024mhs,Yang:2024fkn,Liang:2025hxw}, but the theoretical estimations for these contributions are several orders of magnitude smaller than experiment results \cite{Sheng:2019kmk,Xia:2020tyd}. To explain this discrepancy, it was proposed that the fluctuation of vector meson fields, which effectively describes the strong interaction, is the dominant source for the spin alignment \cite{Sheng:2019kmk,Sheng:2022wsy,Sheng:2022ffb,Sheng:2023urn}. One can also refer to Refs. \cite{Kumar:2022ylt,Kumar:2023ghs,Yang:2024qpy} for discussions on the fluctuation of gluon fields, which also describes the effect of the strong interaction but works at higher temperature than the vector meson field. Assuming that this fluctuation is isotropic in the rest frame of the QGP, the motion of meson relative to the QGP will break the symmetry between its moving direction and transverse directions, leading to stronger transverse fluctuations than the longitudinal fluctuation in the meson's rest frame \cite{Sheng:2023urn}. Consequently, the longitudinally and transversely polarized mesons have different properties and therefore different distributions, resulting in the nontrivial spin alignment. One can also refer to \cite{Sheng:2024kgg,Zhao:2024ipr,Grossi:2024pyh} for discussions on the spectral function of vector meson in a thermal medium, which also provide possible explanations for the spin alignment.   

In experiment, the $\phi$ mesons are reconstructed via their strong decay $\phi\rightarrow K^++K^-$, corresponding to a resonance peak in the invariant mass spectral of the $K^+K^-$ pair. The spin alignment is then related to the polar angle distribution \cite{Schilling:1969um, STAR:2022fan},
\begin{equation}\label{eq:rho00-obs}
\frac{dN_{K^+K^-}}{d\cos\theta^\ast}\propto (1-\rho_{00})+(3\rho_{00}-1)\cos^2\theta^\ast\,,
\end{equation}
where $\theta^\ast$ denotes the polar angle between the daughter $K^+$ meson's momentum direction and the spin quantization direction, observed in the rest frame of the $\phi$ meson, and $N_{K^+K^-}$ is the number of $K^+K^-$ pairs. In this paper, we investigate the production of a $\phi$ meson from a thermal medium and its decay to a $K^+K^-$ pair. The production rate of final-state $K^+K^-$ pairs is related to the propagator and self-energy of the intermediate $\phi$ meson, which will be calculated using the Nambu-Jona-Lasinio (NJL) model \cite{Nambu:1961tp,Nambu:1961fr,Vogl:1991qt,Klevansky:1992qe,Hatsuda:1994pi,Buballa:2003qv,Volkov:2005kw} in this paper. This model serves as a low-energy effective theory for quantum chromodynamics, and has been proved successful in describing the chiral phase transition and meson properties \cite{Lutz:1992dv, Rehberg:1995kh, Buballa:2003qv}.

Within the NJL model, the gluon degrees of freedom are integrated out and reduce to four-fermion interactions. Mesons are usually described as bound states of quark-antiquark pairs, while their propagators are calculated via the random phase approximation with self-energy contributed by a single quark loop \cite{Klevansky:1992qe}. However, such a self-energy is related to processes $s+\bar{s}\rightarrow \phi$ and $\phi\rightarrow s+\bar{s}$, which is unphysical below a critical temperature due to the lack of quark confinement. In experiments, the dominant decay channels of $\phi$ mesons are $\phi\rightarrow K^++K^-$ and $\phi\rightarrow K^0_L+K^0_S$ \cite{ParticleDataGroup:2024cfk}, and therefore a more consistent description of the $\phi$ meson should also include these physical channels. In the framework of the NJL model, they correspond to kaon loop contributions to the $\phi$ meson's self-energy. Compared to the contribution of quark loop, the kaon loop contribution is identified as a next-to-leading order correction in the expansion in $1/N_c$ \cite{Dmitrasinovic:1995cb,Lemmer:1995eb,Oertel:2000sr,Oertel:2000jp,Oertel:1999fk}. However, in this paper we will show that it plays the crucial role at low temperatures, especially for reproducing the $K^+K^-$ invariant mass spectrum in vacuum.

This manuscript is organized as follows. In Sec. \ref{pair-rate} we present a general discussion for the $K^+K^-$ pair production rate through the $\phi$ meson decay and build its relation to the $\phi$ meson's self-energy, the spin alignment of the $\phi$ meson is then extracted in Sec. \ref{sec:spin-alignment}. We present in Sec. \ref{sec:NJL-model} the theoretical framework of the three-flavor NJL model, including the treatments to quarks, mesons, and the effective vertex between them.  Numerical results for the $K^+K^-$ pair production rate in vacuum as well as at finite temperature are present in Sec. \ref{sec:numerical-results} and then the spin alignment of the $\phi$ meson is calculated in Sec. \ref{sec:numerical-spin}. It is worth noting that the propagators for quarks or mesons in this paper differ from those in textbooks by an imaginary unit $i$. For example, the Feynman propagator in this paper is  $\tilde{S}_F(x-y)\equiv -i\left\langle T\psi(x)\overline{\psi}(y)\right\rangle$, which contains an additional $-i$ compared to Ref. \cite{Peskin:1995ev}. The conventions for propagators will not affect our results for properties of quarks and mesons.

\section{$K^{+}K^{-}$ pair production through $\phi$ meson decay} \label{pair-rate}

\begin{figure}[b]
    \centering
    \includegraphics[width=0.5\linewidth]{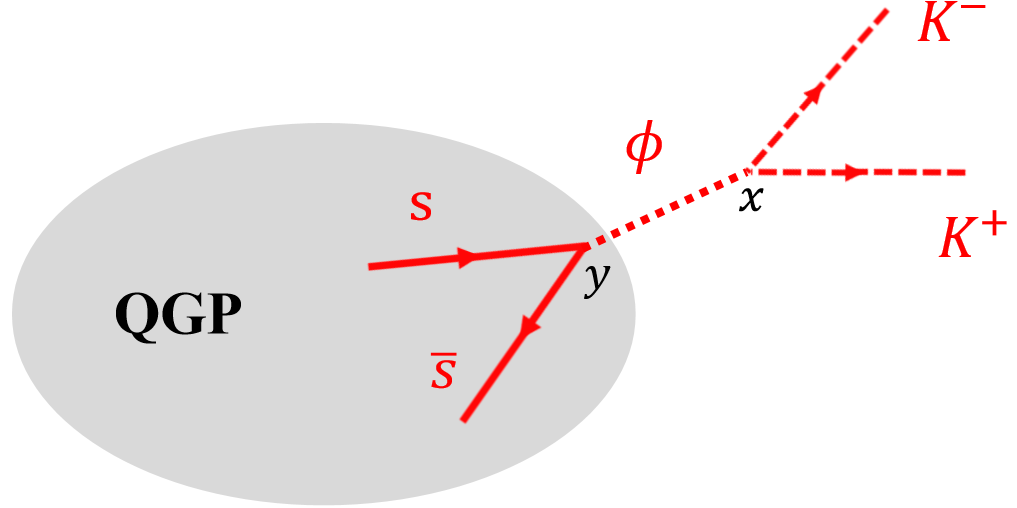}
    \caption{Kaon pair production from the decay of an intermediate $\phi$ meson. The $\phi$ meson is formed at spacetime point $y$ from coalescence of $s$ quark and $\bar{s}$ antiquark, and then decays at spacetime point $x$ to a $K^+K^-$ pair. If the initial state contains kaons, the $\phi$ meson could also be formed through $K^++K^-\rightarrow \phi$.}
    \label{fig:kaon-phi}
\end{figure}

For $\phi$ mesons, the dominant decay channel is $\phi\to K^{+}+K^{-}$ with branch ratio $49.1\%$. In experimental studies on $\phi$ meson's spin alignment, this decay channel is usually considered to reconstruct the $\phi$ mesons. Therefore in this section we will derive the $K^{+}K^{-}$ pair production rate through the decay of $\phi$ mesons. In analogue to the calculation of dilepton production rate \cite{Feinberg:1976ua,McLerran:1984ay,Gale:1990pn,Sheng:2024kgg,Zhao:2024ipr}, we write down the S-matrix element for scattering from an initial state $i$ to a final state $f$ with a $K^+K^-$ pair,
\begin{equation}\label{S-matrix}
    S_{fi}=\int d^{4}x \int d^{4}y\,\langle f,K^{+}K^{-}|J^{K}_{\nu}(x)D_{\text{R,vac}}^{\mu\nu}(x-y)J_{\mu}(y)
|i\rangle\,,
\end{equation}
where $D_{\text{R,vac}}^{\mu\nu}(x-y)$ is the retarded propagator for $\phi$ meson in vacuum. Here $J_{\mu}(y)$ is the source current coupled to the $\phi$ meson, which should be the strangeness quark current if the initial state is the quark-gluon plasma. For the NJL model considered in this paper, the initial system consists of quarks and mesons, thus the source current $J_\mu(y)$ contains both the strangeness quark current and also the kaon current. In Eq. (\ref{S-matrix}), $J_{\nu}^{K}(x)$ is the kaon's current given by
\begin{equation}\label{Kaon-current}
 J_{\nu}^{K}(x)\equiv \int d^{4}x_{1}\int d^{4}x_{2}\,\varphi^{*}(x_{1})\Gamma_{\nu}^\text{vac}(x;x_{1},x_{2})\varphi(x_{2})\,,  
\end{equation}
where $\varphi$ is a complex scalar field denoting kaons, and $\Gamma_{\nu}^\text{vac}(x;x_{1},x_{2})$ is the effective three-meson vertex in vacuum for the process $\phi\to K^{+}+K^{-}$. The phenomenon considered in this process is illustrated by Fig. \ref{fig:kaon-phi}. The quark-antiquark pair (or $K^+K^-$ pair) in the QGP forms a $\phi$ meson at spacetime point $y$. Then the meson propagates from $y$ to another point $x$ and decays to a $K^+K^-$ pair at $x$. The $\phi$ meson is created at the hadronization stage and thus is related to the temperature of the QGP. However, since the lifetime of $\phi$ meson (about 45 fm/c) is much longer than the lifetime of the fireball, we expect that the propagating of $\phi$ meson from $y$ to $x$ and the decay process $\phi\rightarrow K^++K^-$ are weakly affected by the QGP. This is why we use the vacuum propagator and vacuum vertex in Eqs. (\ref{S-matrix}) and (\ref{Kaon-current}). If we study short-lived vector mesons, such as the $\rho$ meson in the $\rho\rightarrow \pi^++\pi^-$ process, we expect that the strong decay happens inside the QGP region and the vacuum propagator and vertex should be replaced by the in-medium ones.   

We consider a box quantization with box volume $V=L^3$ and $L$ being the length of box sides. The kaon field is then quantized as 
\begin{equation}
\varphi(x)=\sum_{\bf p} \frac{1}{\sqrt{2E_{\bf p}V}}\left(a_{\bf p}e^{-ip\cdot x}+b^\dagger_{\bf p} e^{ip\cdot x}\right)\,,
\end{equation}
where ${\bf p}=(2\pi/L)(n_x,n_y,n_z)$ take discrete numbers. In the infinite volume limit, the sum over $n_x,n_y,n_z$ is replaced by a momentum integral, $\lim_{V\rightarrow\infty}\sum_{\bf p}\rightarrow V\int d^3{\bf p}/(2\pi)^3$. Substituting the quantized kaon field into $J^K_\nu(x)$ in Eq. (\ref{Kaon-current}) and then applying $J^K_\nu(x)$ to the state $\left\langle f,K^+K^-\right|$, we obtain 
\begin{equation}
    \langle f, K^{+}K^{-}|J_{\nu}^{K}(x)=\frac{1}{2V\sqrt{E_{+}E_{-}}}\int\frac{d^4q}{(2\pi)^4}\Gamma_{\nu}^\text{vac}(q;p_{+},p_{-})e^{-iq\cdot x}\langle f|\,,
\end{equation}
where $p_\pm$ are momenta for $K^+$ and $K^-$, respectively, with $E_\pm=\sqrt{{\bf p}_\pm^2+m_K^2}$ being their on-shell energies. The effective vertex in momentum space is the Fourier transform of $\Gamma_\nu^\text{vac}(x;x_1,x_2)$,
\begin{equation}
\Gamma_\nu^\text{vac}(q;p_1,p_2)=\int\frac{d^{4}q}{(2\pi)^{4}}\frac{d^4 p_1}{(2\pi)^4}\frac{d^4 p_2}{(2\pi)^4}\,e^{iq\cdot x+ip_1\cdot x_{1}+ip_2\cdot x_{2}}\Gamma_\nu^\text{vac}(x;x_1,x_2)\,.
\end{equation}
The momentum conservation for the decay process ensures that $\Gamma_\nu^\text{vac}(q;p_1,p_2)$ has a non-vanishing value only if $q+p_1+p_2=0$. Thus we can in general extract a delta function for momentum and express the vertex as 
\begin{equation}
\Gamma_\nu^\text{vac}(q;p_1,p_2)=(2\pi)^4\delta^{(4)}(q+p_1+p_2)\tilde{\Gamma}_\nu^\text{vac} (p_1,p_2)\,.
\end{equation}
With the S-matrix (\ref{S-matrix}), we can now evaluate the transition probability per unit spacetime volume,
\begin{eqnarray}
R_{fi}\equiv\lim_{\tau V\rightarrow \infty}\frac{|S_{fi}|^2}{\tau V}
&=&\int\frac{d^3 {\bf p}_+}{(2\pi)^3 2E_+}\frac{d^3 {\bf p}_-}{(2\pi)^3 2E_-}\int d^4y\,e^{-ip\cdot y}\left\langle i \right|J^\ast_{\mu_2}(y/2)\left|f\right\rangle\left\langle f \right|J_{\mu_1}(-y/2)\left|i\right\rangle \nonumber\\
&&\times D_\text{R,vac}^{\mu_1 \nu_1}(p)\tilde{\Gamma}_{\nu_1}^\text{vac}(p_+,p_-)\left[D_\text{R,vac}^{\mu_2 \nu_2}(p)\tilde{\Gamma}_{\nu_2}^\text{vac}(p_+,p_-)\right]^\ast\,,
\end{eqnarray}
where $\tau$ and $V$ denote the interacting time and the volume of the system, and the meson's propagator in momentum space is defined as
\begin{equation}
D_\text{R,vac}^{\mu\nu}(q)\equiv\int\frac{d^{4}q}{(2\pi)^{4}}e^{iq\cdot(x-y)} D_\text{R,vac}^{\mu\nu}(x-y)\,.  
\end{equation}
Assuming that the initial state is a thermal bath, we obtain the production rate of $K^+K^-$ pair by summing over the final state and taking canonical ensemble average over the initial state, 
\begin{equation}
n_{K^+K^-}=\sum_f\sum_i \frac{1}{Z}e^{- E_i/T} R_{fi}\,,
\end{equation}
where $E_i$ denotes the energy of initial state $i$, $T$ is the temperature, and $Z$ is the canonical partition function. Using the spectral function, the result can be expressed as 
\begin{equation}
n_{K^+K^-}=-2\int\frac{d^3 {\bf p}_+}{(2\pi)^3 2E_+}\frac{d^3 {\bf p}_-}{(2\pi)^3 2E_-} n_B(\omega)\tilde{\Gamma}_\mu^{\text{vac}\,\ast}(p_+,p_-)\rho ^{\mu\nu}(p)\tilde{\Gamma}_\nu^\text{vac}(p_+,p_-)\,,
\end{equation}
where $p^\mu=p_+^\mu +p_-^\mu$ is the total momentum of $K^+K^-$ pair, $\omega=E_++E_-$ is the total energy, and $n_B(\omega)=1/[\exp(\omega/T)-1]$ is the Bose-Einstein distribution. The spectral function $\rho ^{\mu\nu}(p)$ is related to the imaginary part of the $\phi$ meson‘s self-energy as
\begin{equation}\label{eq:spectral-function}
\rho^{\mu\nu}(p)=-\left[D_\text{R,vac}^{\alpha\mu}(p)\right]^{*}\left[\mathrm{Im} \Pi^\text{R}_{\alpha\beta}(p)\right] D_\text{R,vac}^{\beta\nu}(p)\,,
\end{equation}
where the retarded self-energy is defined by a current-current correlation,
\begin{equation}\label{eq:self-energy}
\Pi^\text{R}_{\mu\nu}(p)\equiv -i\int d^4y\,\theta(y^0) e^{-ip\cdot y}\left\langle \left[J_\mu(y),J_\nu(0)\right]\right\rangle\,.
\end{equation}

\section{Relating $K^+K^-$ pair production to $\phi$ meson's spin alignment} \label{sec:spin-alignment}

In general, the momenta ${\bf p}_\pm$ for $K^\pm$ can be described by $\{p^\mu,\,\theta^\ast,\,\phi^\ast\}$, where $p^\mu$ is the total momentum of the $K^+K^-$ pair, $\theta^\ast$ and $\phi^\ast$ are the polar angle and the azimuthal angle of $K^+$'s momentum relative to a specified direction in the rest frame of the $K^+K^-$ pair. Then the momentum integral measure $d^3{\bf p}_+ d^3{\bf p}_-$ is replaced by $|\mathcal{J}|d^4p d\theta^\ast d\phi^\ast$, where $|\mathcal{J}|$ is the determinant for the Jaccobi matrix for the variable substitution. An explicit calculation shows that
\begin{equation}
d^3{\bf p}_+ d^3{\bf p}_-=\frac{1}{2}E_+E_-\sqrt{1-\frac{4M_K^2}{M_\phi^2}}d^4p\,\sin\theta^\ast d\theta^\ast d\phi^\ast\,,
\end{equation}
where $\omega$ is the energy and $M_\phi=\sqrt{\omega^2-{\bf p}^2}$ is the invariant mass of the $K^+K^-$ pair. Then the differential production rate reads
\begin{equation}\label{eq:KK-production-rate}
    \frac{dn_{K^+K^-}}{d^{4}p\,d\cos\theta^{*} d\phi^{*}}=-\frac{1}{4(2\pi)^6}\sqrt{1-\frac{4M^{2}_{K}}{M_\phi^{2}}}n_{B}(\omega)\tilde{\Gamma}_{\mu}^{\text{vac}\,\ast}(p_{+},p_{-})\rho^{\mu\nu}(p)\tilde{\Gamma}_{\nu}^\text{vac}(p_{+},p_{-})\,. 
\end{equation}
As spin-1 particles, the $\phi$ mesons have three spin states $\lambda=0,\pm1$. Therefore the spectral function can be projected to the $3\times 3$ spin space as 
\begin{equation}\label{eq:propagator-decomposition}
\rho^{\mu\nu}(p)=-\sum_{\lambda,\lambda^\prime=0,\pm 1}\epsilon^\ast_\mu(\lambda,p)\epsilon_\nu(\lambda^\prime,p)\xi_{\lambda\lambda^\prime}(p)\,,
\end{equation}
where the spin polarization vectors are given by 
\begin{equation}\label{eq:Polarization-vector}
\epsilon^\mu(\lambda,p)=\left(\frac{{\bf p}\cdot\boldsymbol{\epsilon}_\lambda}{M_\phi},\boldsymbol{\epsilon}_\lambda+\frac{{\bf p}\cdot\boldsymbol{\epsilon}_\lambda}{M_\phi(\omega+M_\phi)}{\bf p}\right)\,.
\end{equation}
Here $\boldsymbol{\epsilon}_\lambda$, $\lambda=0,\pm1$, are 3-component vectors which are properly normalized as $\boldsymbol{\epsilon}_\lambda^\ast\cdot\boldsymbol{\epsilon}_{\lambda^\prime}=\delta_{\lambda\lambda^\prime}$. It is easy to check that these vectors are perpendicular to $p^\mu$, $p_\mu\epsilon^\mu(\lambda,p)=0$, ensuring $p_\mu\rho^{\mu\nu}(p)=0$. These vectors satisfy $\boldsymbol{\epsilon}_0^\ast=\boldsymbol{\epsilon}_0$ and $\boldsymbol{\epsilon}_{\pm 1}^\ast=-\boldsymbol{\epsilon}_{\mp 1}$, while the direction of $\boldsymbol{\epsilon}_0$ is also known as the spin quantization direction. In experiments, the most widely used choice is the event plane direction $\boldsymbol{\epsilon}_0=(0,1,0)$ or the momentum direction $\boldsymbol{\epsilon}_0={\bf p}/|{\bf p}|$. 
In this paper, we consider the motion of $\phi$ meson relative to a thermal background, which breaks the symmetry between the direction of motion and the perpendicular directions. As a consequence, it is convenient to choose the spin quantization direction as the momentum direction. Then the spin polarization vector in the rest frame of $\phi$ meson with magnetic quantum number $\lambda=0$ ($\lambda=\pm1$) is parallel (perpendicular) to the direction of ${\bf p}$, which will be referred as longitudinally (transversely) polarized in later discussions. In this case, the spin quantization direction is 
\begin{equation}
\epsilon^\mu_\text{H}=\frac{1}{M_\phi}\left(|{\bf p}|,\,\omega \frac{{\bf p}}{|{\bf p}|}\right)\,,
\end{equation}
corresponding to the so-called {\it Helicity Frame} \cite{Schilling:1969um,Faccioli:2010kd}. 
In this frame, $\xi_{\lambda\lambda^\prime}$ in Eq. (\ref{eq:propagator-decomposition}) is diagonal in spin space and the spectral function can be decomposed into longitudinal and transverse components as
\begin{equation}\label{eq:decomposed-propagator}
\rho^{\mu\nu}(p)=-\epsilon^\mu_\text{H}(p)\epsilon^\nu_\text{H}(p)\rho_{L}(p)+\left[g^{\mu\nu}-\frac{p^\mu p^\nu}{p^2}+\epsilon^\mu_\text{H}(p)\epsilon^\nu_\text{H}(p)\right]\rho_{T}(p)\,.
\end{equation}
Substituting Eq. (\ref{eq:decomposed-propagator}) into Eq. (\ref{eq:KK-production-rate}), the differential production rate of $K^+K^-$ pairs is now expressed as 
\begin{eqnarray}\label{eq:KK-production-rate-new}
    \frac{dn_{K^+K^-}}{d^{4}p\,d\cos\theta^{*} d\phi^{*}}&=&\frac{1}{4(2\pi)^6}\sqrt{1-\frac{4M^{2}_{K}}{M_\phi^{2}}}n_{B}(\omega) \nonumber \left\{\left|\epsilon^\mu_\text{H}(p)\tilde{\Gamma}_{\mu}^\text{vac}(p_{+},p_{-})\right|^2 \left[\rho_{L}(p)-\rho_{T}(p)\right]\right.\nonumber \\
    &&\left.-\tilde{\Gamma}_{\mu}^{\text{vac}\,\ast}(p_{+},p_{-})\tilde{\Gamma}_{\nu}^\text{vac}(p_{+},p_{-})\left(g^{\mu\nu}-\frac{p^\mu p^\nu}{p^2}\right) \rho_{T}(p)\right\}\,,
\end{eqnarray}
where $\rho_{L,T}$ represent spectral functions for longitudinally and transversely polarized modes, respectively,
\begin{eqnarray}\label{eq:spectral-functions-LT}
\rho_L(p)&=&-\epsilon_\mu^\text{H}(p)\epsilon_\nu^\text{H}(p)\rho^{\mu\nu}(p)\,,\nonumber\\
\rho_T(p)&=&\frac{1}{2}\left[g_{\mu\nu}-\frac{p_\mu p_\nu}{p^2}+\epsilon_\mu^\text{H}(p)\epsilon_\nu^\text{H}(p)\right]\rho^{\mu\nu}(p)\,,
\end{eqnarray}
with $\rho^{\mu\nu}$ being related to the $\phi$ meson's self-energy and propagator as shown in Eqs. (\ref{eq:spectral-function}) and (\ref{eq:self-energy}).

In experiments, the $\phi$ meson's spin alignment is measured through the polar angle distribution of $K^+K^-$ pairs as shown in Eq. (\ref{eq:rho00-obs}).
Comparing Eq. (\ref{eq:KK-production-rate-new}) to Eq. (\ref{eq:rho00-obs}) we can then extract the spin alignment $\rho_{00}$ as a function of momentum $p^\mu$. In the vacuum case, the effective vertex $\tilde{\Gamma}_\mu^\text{vac}(p_+,p_-)$ must be a linear combination of the relative momentum $q^\mu\equiv p_+^\mu-p_-^\mu$ and the total momentum $p^\mu$. However, the part linear in $p^\mu$ does not contribute in Eq. (\ref{eq:KK-production-rate-new}) because of $p_\mu\epsilon^\mu_\text{H}=0$ and $p_\mu(g^{\mu\nu}-p^\mu p^\nu/p^2)=0$. Therefore Eq. (\ref{eq:KK-production-rate-new}) reduces to the following form,
\begin{equation}\label{eq:pair-production-DR-TL}
\frac{dn_{K^+K^-}}{d^{4}p\,d\cos\theta^{*} d\phi^{*}}\propto \left[q_\mu\epsilon^\mu_\text{H}(p)\right]^2\left[\rho_{L}(p)-\rho_{T}(p)\right] -q^2 \rho_{T}(p)\,.
\end{equation}
Supposing that the produced kaons are on the mass-shell, we can prove that $\left[q_\mu\epsilon^\mu_\text{H}(p)\right]^2=(M_\phi^2-4M_K^2)\cos^2\theta^\ast$ and $q^2=4M_K^2-M_\phi^2$, and therefore the spin alignment reads,
\begin{equation}
\rho_{00}(p)=\frac{\rho_{L}(p)}{\rho_{L}(p)+2\,\rho_{T}(p)}\,,
\end{equation}
as expected. A detailed calculation for the $K^+K^-$ pair production rate and the spin alignment will be given in Sec. \ref{sec:numerical-results} in the framework of the NJL model.  

\section{$SU(3)$ NJL model} \label{sec:NJL-model}

\subsection{General framework}

As shown by Eqs. (\ref{eq:spectral-function}), (\ref{eq:self-energy}), (\ref{eq:KK-production-rate-new}), and (\ref{eq:spectral-functions-LT}), the calculation of $K^+K^-$ pair production rate needs the explicit expression of the $\phi$ meson's self-energy and propagator. In this paper, we implement the $SU(3)$ NJL model to describe quarks, mesons, and interactions between them.  
The Lagrangian density reads \cite{Bernard:1987sg,Klimt:1989pm,Vogl:1991qt,Hatsuda:1994pi},
\begin{align}
\mathcal{L} =
  &\,\bar{\psi}(i\gamma_{\mu}\partial^{\mu}-m)\psi +G_{S}\sum_{a=0}^{8}\Big[(\bar{\psi}\lambda_{a}\psi)^{2}+(\bar{\psi}i\gamma_{5}\lambda_{a}\psi)^{2}\Big]\nonumber \\
 & +G_{V}\sum_{a=0}^{8}\Big[(\bar{\psi}\gamma_{\mu}\lambda_{a}\psi)^{2}+(\bar{\psi}i\gamma_{\mu}\gamma_{5}\lambda_{a}\psi)^{2}\Big]\nonumber \\
 &-K\Big[\text{det}\,\bar{\psi}(1+\gamma_{5})\psi+\text{det}\,\bar{\psi}(1-\gamma_{5})\psi\Big]\,,
\end{align}
where $\psi=(\psi_u,\,\psi_d,\,\psi_s)^T$ is a column vector of Dirac spinors, $m=\text{diag}\{m_u,\,m_d,\,m_s\}$ is a diagonal matrix in flavor space with diagonal elements being bare masses for $u$, $d$, and $s$ quarks, respectively, and $\lambda_a$, $a=0,1,\cdots,8$, are $3\times 3$ Gell-Mann matrices. Here we have included scalar and vector four-fermion interactions as well as the six-fermion KMT interaction, with $G_S$, $G_V$, and $K$ represent the corresponding coupling constants. Under the mean-field approximation, the Lagrangian is converted to 
\begin{equation}
\mathcal{L}_\text{MF}=\sum_{f=u,d,s}\bar{\psi}_{f}(i\gamma_{\mu}\partial^{\mu}-M_{f})\psi_{f}-2G_{S}\sum_{f=u,d,s}\sigma_{f}^{2}+4K\sigma_{u}\sigma_{d}\sigma_{s}\,,
\end{equation}
where we have included the chiral condensate $\sigma_f\equiv\langle\bar\psi_f\psi_f\rangle$. The quark's dynamic mass $M_f$ is defined by
\begin{equation}
M_f=m_f-4G_S\sigma_f+2K\prod_{f^\prime\neq f}\sigma_{f^\prime}\,.
\end{equation}
For a thermal equilibrium system at finite temperature $T$, the total thermodynamic potential is given by 
\begin{equation}
\Omega_\text{MF}=\sum_{f} \left(2G_S\sigma_f^2-\Omega_f\right)-4K\sigma_u\sigma_d\sigma_s\,,
\end{equation}
where the contribution from quark with flavor $f=u,d,s,$ reads,
\begin{equation}\label{eq:thermodynamic-potential}
    \Omega_f=2N_c\int\frac{d^3{\bf p}}{(2\pi)^3}\left[E_f+2T\ln(1+e^{-E_f/T})\right]\,.
\end{equation}
Here the quark's on-shell energy $E_f\equiv\sqrt{{\bf p}^2+M_f^2}$ depends on its dynamic mass $M_f$. By minimizing $\Omega_\text{MF}$ with respect to chiral condensates, i.e., $\partial\Omega_\text{MF}/\partial \sigma_f=0$ for $f=u,d,s$, one can evaluate $\sigma_f$ and consequently obtain the dynamic masses.

Since the NJL model is is non-renormalizable, we need to employ a regularization scheme to avoid ultraviolet divergences in momentum integrals. 
For the thermodynamic potential (\ref{eq:thermodynamic-potential}) and other loop integrals in later discussions, we separate them into the vacuum parts, which are independent to temperature, and the medium parts, which rely on the temperature through the fermi-Dirac distribution or $\ln(1+e^{-E_f/T})$. The medium parts vanish at zero temperature and do not need any regularization because high-momentum contributions are automatically suppressed by thermal distributions. For the vacuum parts, we apply a hard cutoff for integral variables. In later discussions we will calculate self-energies of kaons and $\phi$ mesons as well as the $\phi K^+K^-$ vertex for mesons with nonvanishing 3-momentum ${\bf p}$. In order to ensure the covariance of these quantities, we always perform the hard cutoff in the local rest frame of the considered meson. The self-energy or the vertex in the frame where the meson has nonzero 3-momentum is then obtained by performing the Lorentz boost. 
We take quark bare masses $m_u=m_d=5.5$ MeV, $m_s=140.7$ MeV, the cutoff of momentum $\Lambda=602.3$ MeV, $G_S=5.058\text{ GeV}^{-2}$, and $K=155.9\text{ GeV}^{-5}$, as given in \cite{Rehberg:1995kh}, which are determined by fitting physical masses and decay constants of mesons in vacuum. Here the cutoff parameter $\Lambda$ applies to all quark loop integrals in this work. On the other hand, when evaluating the $\phi$ meson's self-energy, we go beyond leading order in the $1/N_c$ expansion by including the meson-loop contributions. This meson loop is dynamically independent of the quark sector and thus could be regularized independently to the quark loop by introducing another hard cutoff parameter $\Lambda_M\neq\Lambda$. This parameter and the vector coupling $G_V$ are determined by comparing the invariant mass spectrum of $K^+K^-$ pair to the one observed in experiment \cite{STAR:2022fan}. The results reads $G_V = -1.0621~\text{GeV}^{-2}$ and $\Lambda_M = 0.719$ GeV.

\begin{figure}[t]
    \centering
    \includegraphics[width=0.5\linewidth]{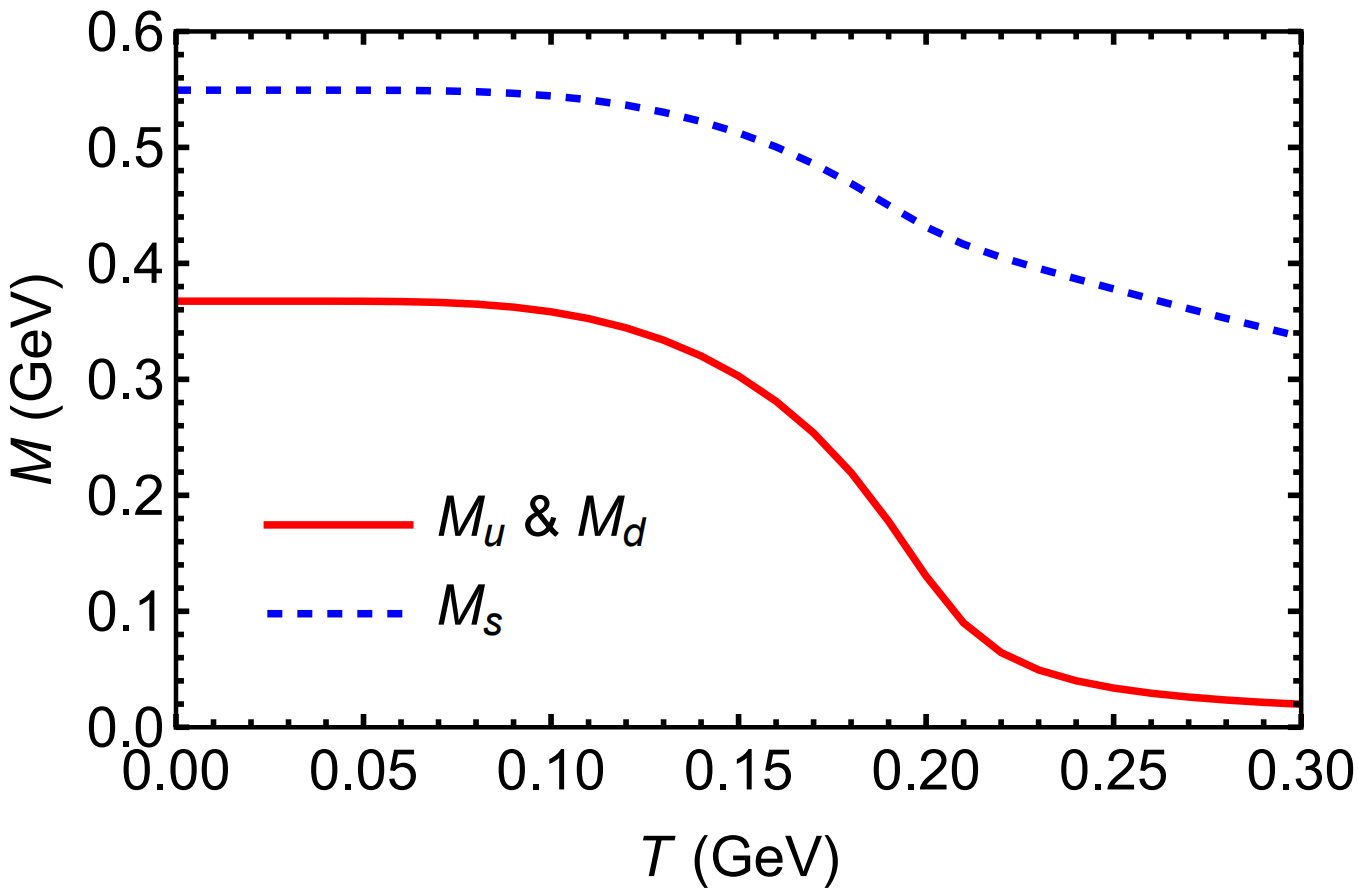}
    \caption{The dynamical masses for light quarks ($u$ and $d$) and the $s$ quark as functions of the temperature.}
    \label{fig:Quark-dynamic-mass}
\end{figure}

We show in Fig. \ref{fig:Quark-dynamic-mass} the dynamic masses for $u$, $d$, and $s$ quarks as functions of the temperature. Due to the isospin symmetry, $u$ and $d$ quarks have the same mass, which is lower than the mass of the $s$ quark. As the temperature increases, the dynamic masses decrease as indicated by the crossover of chiral phase transition.  

\subsection{Kaon's mass and coupling with quarks}

\begin{figure}[t]
    \centering
    \includegraphics[width=0.47\linewidth]{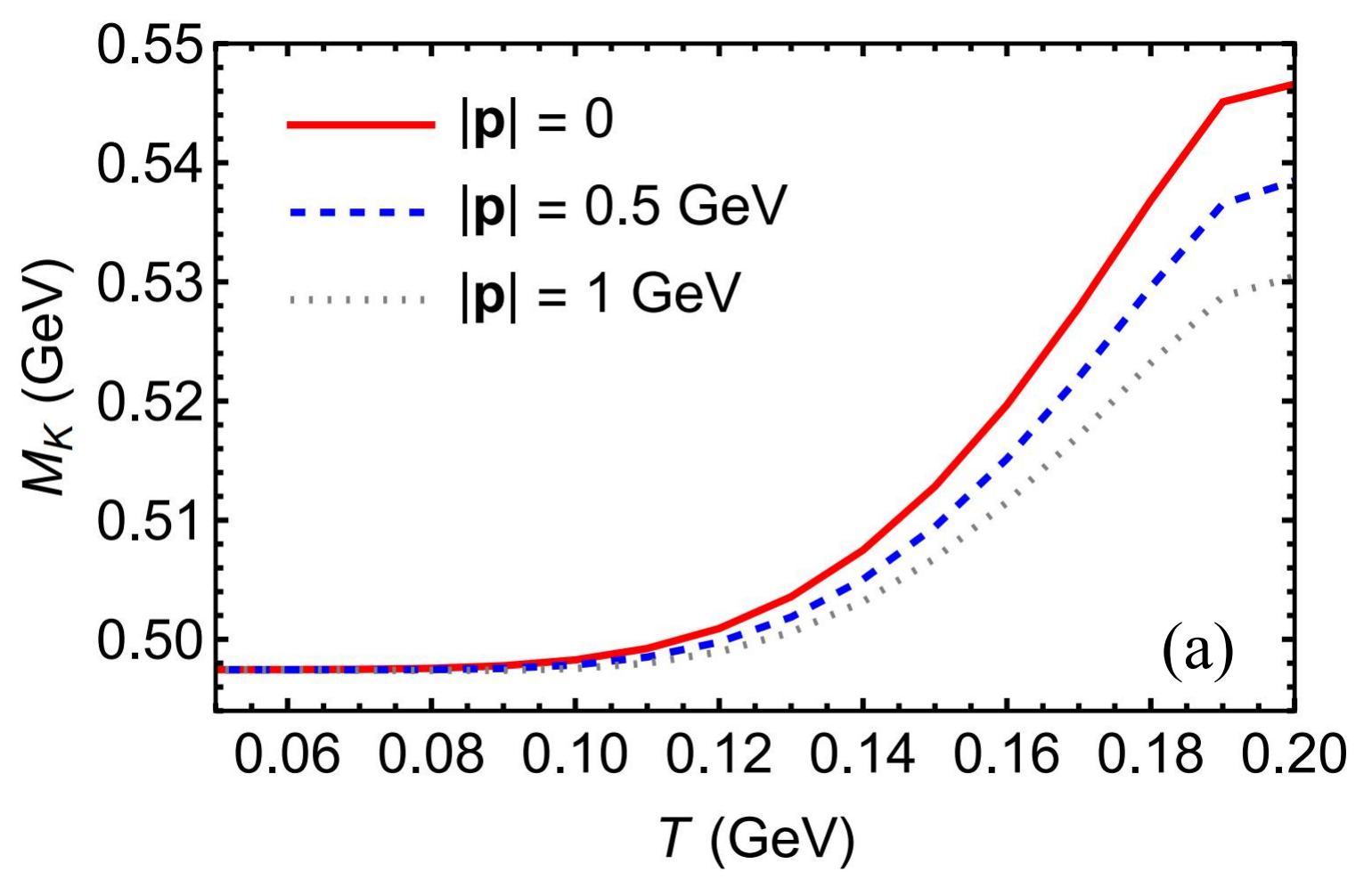}
    \includegraphics[width=0.45\linewidth]{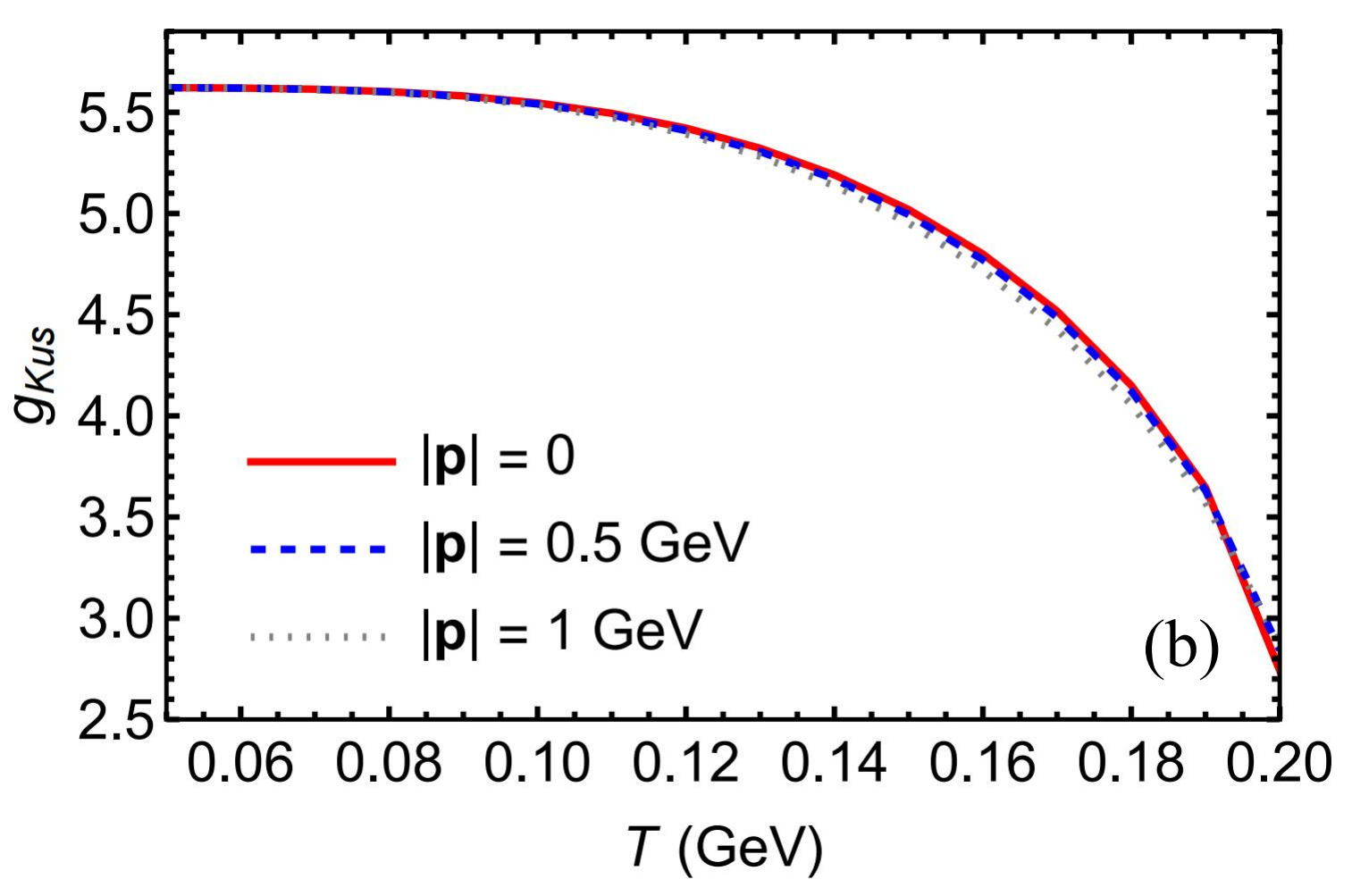}
    \caption{The dynamic mass of $K^\pm$ (left panel) and the effective coupling strength $g_{Kus}$ (right panel) as functions of temperature $T$ for kaon with momentum $|{\bf p}|$=0 (red lines), 0.5 GeV (blue dashed lines), and $1$ GeV (gray dotted lines).}
    \label{fig:kaon-mass-coupling}
\end{figure}

We now focus on the $K^\pm$ mesons in the framework of the NJL model. The propagator of kaon obeys a Dyson-Schwinger equation, which has the following solution,
\begin{equation}\label{eq:Kaon-propagator}
D_K(p)=\frac{4G_4^+}{1-4G_4^+\Pi_{K}(p)}\,,
\end{equation}
where the pseudoscalar coupling is $G_4^+=G_S-K\sigma_d/2$, and the self-energy of kaon reads,
\begin{equation}
\Pi_K(p)\equiv -iN_c\int\frac{d^4q}{(2\pi)^4}\text{Tr}\left[\gamma_5 \tilde{S}_u(q)\gamma_5 \tilde{S}_s(q-p)\right]\,.
\end{equation}
Here $\tilde{S}_u$ and $\tilde{S}_s$ denote propagators of $u$ and $s$ quarks, respectively, which are assumed to have the following on-shell form,
\begin{equation}\label{eq:quark-propagator}
\tilde{S}_f(q)=\frac{\gamma\cdot q +M_f}{q^2-M_f^2}\,.
\end{equation}
Below a critical temperature, kaons can exist as stable particles with pole mass determined by the gap equation $1-4G_4^+\Pi_K(p)=0$. Then under the dipole approximation, the propagator in  Eq. (\ref{eq:Kaon-propagator}) can be approximate as
\begin{equation} \label{eq:dipole-approximation}
 D_K(p)\approx \frac{g_{Kus}^2(p)}{q^2-M_K^2(p)}\,,
\end{equation}
where the effective coupling strength between kaon and $u$, $s$ quarks is determined by 
\begin{equation}
g_{Kus}(p)=\left[\frac{\partial \Pi_K(p)}{\partial p^2}\bigg|_{p^2=M_K^2}\right]^{-1/2}\,.
\end{equation}
We note that at finite temperature, the kaon's dynamic mass $M_K$ and coupling strength $g_{Kus}$ are all functions of the kaon's momentum relative to the thermal background. Figure \ref{fig:kaon-mass-coupling} presents $M_K$ (left panel) and $g_{Kus}$ (right panel) as functions of temperature for kaons with momenta $|{\bf p}|= 0$, 0.5, and $1$ GeV. In these figures, we concentrate on the temperature interval $0.05$ GeV $<T<$ 0.2 GeV. For $T<0.05$ GeV, $M_K$ and $g_{Kus}$ are nearly independent to the temperature. Numerical calculations show that the variations of $M_K$ and $g_{Kus}$ at $T<0.05$ GeV are $\delta M_K\sim 10^{-5}$ GeV and $\delta g_{Kus}\sim 10^{-3}$, which are sufficiently small compared to vacuum values of $M_K$ and $g_{Kus}$. This is why properties at $T<0.05$ GeV are not shown here.  Above 0.2 GeV, we cannot find a real root for the gap equation of kaon, indicating that Eq. (\ref{eq:dipole-approximation}) is no longer a good approximation.  We observe from Fig. \ref{fig:kaon-mass-coupling} that $M_K$ increases, while $g_{Kus}$ decreases, for increasing temperature. Kaons with larger momenta have smaller masses but $g_{Kus}$ is nearly independent to $|{\bf p}|$. 

\subsection{Self-energy and propagator of $\phi$ meson}

\begin{figure}[b]
    \centering
    \includegraphics[width=0.4\linewidth]{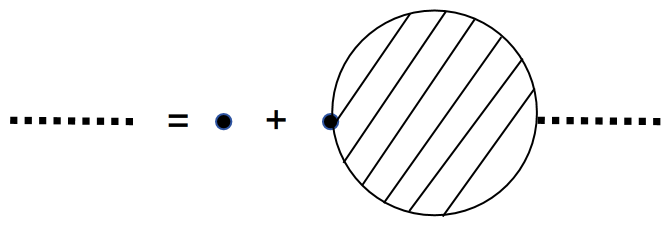}
    \caption{Dyson-Schwinger equation for vector meson. Here dotted lines denote full propagators of meson, black dots denote the four-fermion vertex, and shaded circle denotes the meson's self-energy.}
    \label{fig:D-S-equation}
\end{figure}

\begin{figure}[t]
    \centering
    \includegraphics[width=0.8\linewidth]{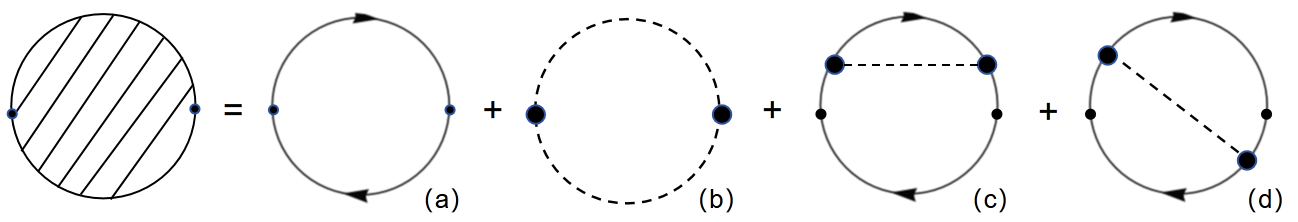}
    \caption{Contributions to the self-energy of the $\phi$ meson: (a) denotes leading order contribution from a $s$-quark loop, (b) the contribution from a kaon loop, and (c), (d) corrections at the same order as (b). Solid lines with arrows indicate the quark propagators, and dashed lines the propagators of kaons. }
    \label{fig:self-energy}
\end{figure}

For a vector meson in the NJL model, the propagator obeys the Dyson-Schwinger equation, as illustrated in Fig. \ref{fig:D-S-equation},
\begin{equation}\label{eq:D-S-equation}
D^{\mu\nu}(p)=4G_{V}\Delta^{\mu\nu}(p)+4G_{V}\Delta^{\mu\alpha}(p)\Pi^\text{tot}_{\alpha\beta}(p)D^{\beta\nu}(p)\,,
\end{equation}
where $D^{\mu\nu}(p)$ denotes the full propagator of a meson with momentum $p^\mu$, $\Delta^{\mu\nu}\equiv g^{\mu\nu}-p^\mu p^\nu/p^2$ is the projection operator ensuring that the propagator is perpendicular to momentum, i.e., $p_\mu D^{\mu\nu}(p)=0$, and $\Pi^\text{tot}_{\alpha\beta}$ is the meson's total self-energy. In this paper, we focus on the $\phi$ meson, which consists of $s$ and $\bar{s}$. Therefore the leading order contribution to the self-energy is from a $s$ quark loop. However, this corresponds to a decay channel $\phi\rightarrow s+\bar{s}$, which is unphysical at low temperatures. In experiments, $\phi$ mesons can decay to $K^+K^-$ or $K^0_LK^0_S$ pairs with branch ratios $49.1\%$ and $33.9\%$, respectively. To address these physical decay channels, we introduce contributions of kaon loops to the $\phi$ meson's self-energy. The total self-energy is illustrated by Fig. \ref{fig:self-energy},
\begin{equation}\label{eq:self-energy-tot}
\Pi^\text{tot}_{\mu\nu}=\Pi^\text{Q-loop}_{\mu\nu}+\Pi^\text{K-loop}_{\mu\nu}+\Pi^\text{K-tad}_{\mu\nu}\,,
\end{equation}
where $\Pi^\text{Q-loop}_{\mu\nu}$ and $\Pi^\text{K-loop}_{\mu\nu}$ are contributions from the quark loop and the kaon loop, respectively, as shown by Fig. \ref{fig:self-energy} (a) and (b). The kaon loop is incorporated in the self-energy as a higher order correction in $1/N_c$ relative to the quark loop, with $N_c=3$ denoting the number of colors. We also include $\Pi^\text{K-tad}_{\mu\nu}$ to denote contributions from Fig. \ref{fig:self-energy} (c) and (d), which is the same order as $\Pi^\text{K-loop}_{\mu\nu}$. We emphasize that the model in this paper keeps the isospin symmetry between $u$ and $d$ quarks, indicating that $K^\pm$, $K_L^0$, and $K_S^0$ have identical properties except their charges. For the $\phi$ meson's self-energy, the contribution of a $K^+K^-$ loop is exactly the same as that of a $K^0_LK^0_S$ loop.

Using propagators of quark and kaon, we express $\Pi^\text{Q-loop}_{\mu\nu}$ and $\Pi^\text{K-loop}_{\mu\nu}$ as follows,
\begin{eqnarray}\label{self-energy}
\Pi^\text{Q-loop}_{\mu\nu}(p)&=&iN_c\int\frac{d^4 q}{(2\pi)^4}\text{Tr}\left[\gamma_\mu \tilde{S}_s(q)\gamma_\nu \tilde{S}_s(q-p)\right]\,, \nonumber\\
\Pi^\text{K-loop}_{\mu\nu}(p)&=&-i N_K f_{\phi KK}^2\int\frac{d^4 q}{(2\pi)^4}\tilde\Gamma_\mu(q-p,-q)\left[\tilde\Gamma_\nu(q-p,-q)\right]^\ast \tilde{D}_K(q) \tilde{D}_K(q-p)\,,
\end{eqnarray}
where $N_K=2$ is because we have considered two identical decay channels, $\phi\rightarrow K^++ K^-$ and $\phi\rightarrow K_L^0+K_S^0$, and $f_{\phi KK}=2$ accounts for the flavor factor from Gell-Mann matrices \cite{Rehberg:1995kh}. Here $\tilde{S}_s(q)$ is the propagator of $s$ quark, given in Eq. (\ref{eq:quark-propagator}), while $\tilde{D}_K(q)=1/(q^2-M_K^2)$ is the propagator of kaon.
For the contribution of $\Pi^\text{K-tad}_{\mu\nu}$, we do not perform explicit calculations but simply assume that $\Pi^\text{K-tad}_{\mu\nu}\propto g_{\mu\nu}$ and impose the constraint $p^\mu\Pi^\text{tot}_{\mu\nu}(p)=0$. It is straightforward to obtain 
\begin{equation}\label{eq:self-energy-tad}
\Pi^\text{K-tad}_{\mu\nu}(p)=-g_{\mu\nu}\frac{p^\alpha p^\beta}{p^2}\Pi_{\alpha\beta}^\text{K-loop}(p)\,.
\end{equation}
Such an assumption has been widely used in literature and has been verified by direct calculations \cite{Lemmer:1995eb,He:1997gn}. 

We focus on moving $\phi$ mesons in a thermal medium. In the helicity frame, the $\phi$ meson's propagator and self-energy are decomposed into longitudinal and transverse components in a similar way as Eq. (\ref{eq:decomposed-propagator}). For the Dyson-Schwinger equation in (\ref{eq:D-S-equation}), we derive the following solutions, 
\begin{equation}
    D_{L/T}(p)=\frac{4G_{V}}{1+4G_{V}\Pi^\text{tot}_{L/T}(p)}.
\end{equation}
With the help of Eqs. (\ref{eq:spectral-function}) and (\ref{eq:spectral-functions-LT}), we obtain spectral functions for longitudinally and transversely polarized modes, 
\begin{equation}\label{eq:spectral-function-new}
\rho_{L/T}(p)=-\left|\frac{4G_V}{1+4G_V\Pi^\text{tot}_\text{vac}(p)}\right|^2\,\text{Im}\,\Pi_{L/T}^\text{tot}(p)\,,
\end{equation}
where $\Pi^\text{tot}_\text{vac}(p)$ is the self-energy in vacuum, i.e., at zero temperature. Since the vacuum is Lorentz invariant, the longitudinally and transversely polarized modes has the same self-energy $\Pi^\text{tot}_\text{vac}(p)=\lim_{T\rightarrow0}\Pi_{L/T}^\text{tot}(p)$. However, at finite temperature, $\Pi_{L}^\text{tot}$ and $\Pi_{T}^\text{tot}$ could have a finite difference, leading to a nontrivial spin alignment.

\subsection{$\phi K^{+}K^{-}$ vertex}

\begin{figure}[t]
    \centering
    \includegraphics[width=0.4\linewidth]{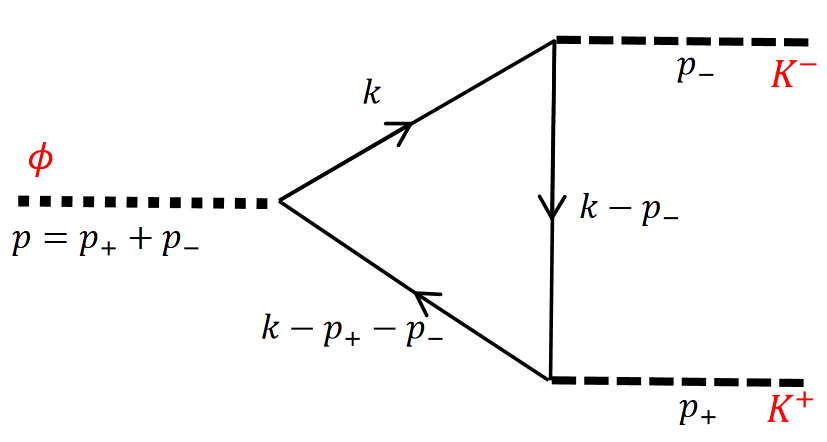}
    \caption{Effective vertex between $\phi$ and $K^\pm$ mesons. Solid lines, dashed lines, and the dotted line denote the propagator of quark, the outgoing kaons, and the incoming $\phi$ meson, respectively.}
    \label{fig:triangle-diagram}
\end{figure}

For calculations of the decay process $\phi \to K^{+}+K^{-}$ and the kaon loop contribution to the $\phi$ meson's self-energy, the $\phi K^{+}K^{-}$ vertex is an essential element. In the NJL model, this vertex is described by a triangle quark loop, as shown in Fig. \ref{fig:triangle-diagram}. Using Feynman rules, we can write down the effective vertex,
\begin{equation}\label{eq:triangle-vertex}
\tilde{\Gamma}^\mu(p_+,p_-)=-N_cg_{Kus}(p_+)g_{Kus}(p_-)\int\frac{d^4 k}{(2\pi)^4}\text{Tr}\left[\gamma^\mu \tilde{S}_s (k-p_+-p_-)\gamma^5\tilde{S}_u(k-p_-)\gamma^5\tilde{S}_s(k)\right]\,,
\end{equation}
where $p_\pm$ denotes momenta of outgoing $K^\pm$ and the momentum of incoming $\phi$ meson is $p=p_++p_-$. The factors $g_{Kus}$ are coupling strength between $K^\pm$ and $u$, $s$ quarks. In principle, we should also take the coupling between $\phi$ meson and $s$ quarks into account. In this paper, however, such a coupling is included in the propagator of $\phi$ meson and thus does not appear in Eq. (\ref{eq:triangle-vertex}).

In the vacuum case, the vertex $\tilde\Gamma^\mu_\text{vac}(p_+,p_-)$ can be expressed as a linear combination of Lorentz vectors $p^\mu=p_+^\mu+p_-^\mu$ and $q^\mu=p_+^\mu-p_-^\mu$,
\begin{equation}
\tilde{\Gamma}^\mu_\text{vac}(p_+,p_-)=q^\mu\Gamma_{1,\text{vac}}(p_+,p_-)+p^\mu\Gamma_{2,\text{vac}}(p_+,p_-)\,.
\end{equation}
At finite temperature, the decomposition of $\tilde\Gamma^\mu$ becomes much more complicated because of the thermal background. In general, we have
\begin{equation}\label{Gamma-decomposition}
\tilde{\Gamma}^\mu(p_+,p_-)=q^\mu\Gamma_{1}(p_+,p_-)+p^\mu\Gamma_{2}(p_+,p_-)+(u\cdot p)u^\mu\Gamma_3(p_+,p_-)+\frac{1}{u\cdot p}\epsilon^{\mu\nu\alpha\beta}q_\nu p_\alpha u_\beta \Gamma_4(p_+,p_-)\,.
\end{equation}
where $u^\mu=(1,0,0,0)$ denotes the rest frame of the thermal background. The coefficients can be evaluated numerically using Eq. (\ref{eq:triangle-vertex}), whose properties will be shown in the upcoming subsection.

\subsection{Contribution from kaon loop to $\phi$ meson's self-energy}

\begin{figure}[t]
    \centering
    \includegraphics[width=0.446\linewidth]{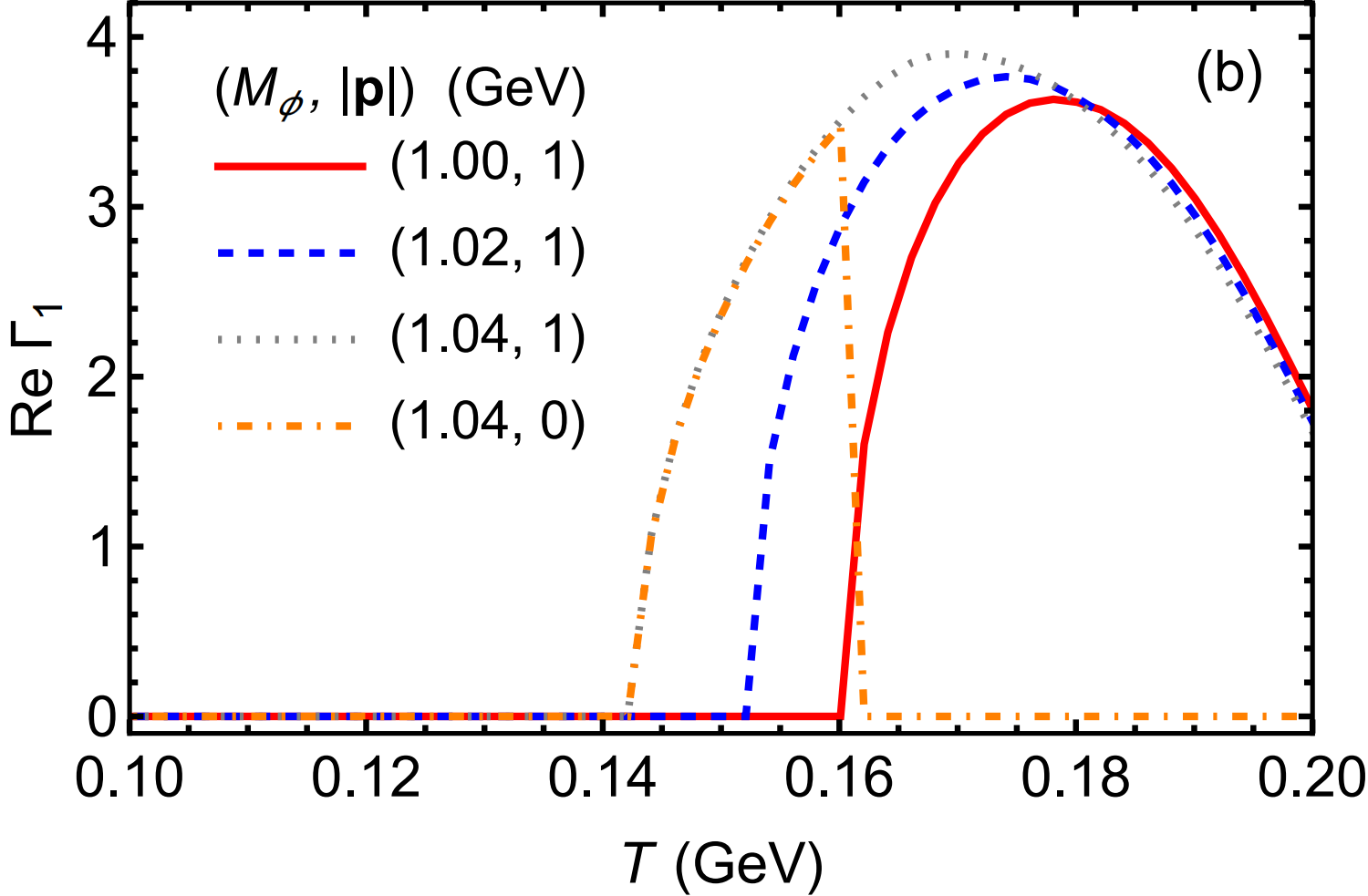}
    \includegraphics[width=0.45\linewidth]{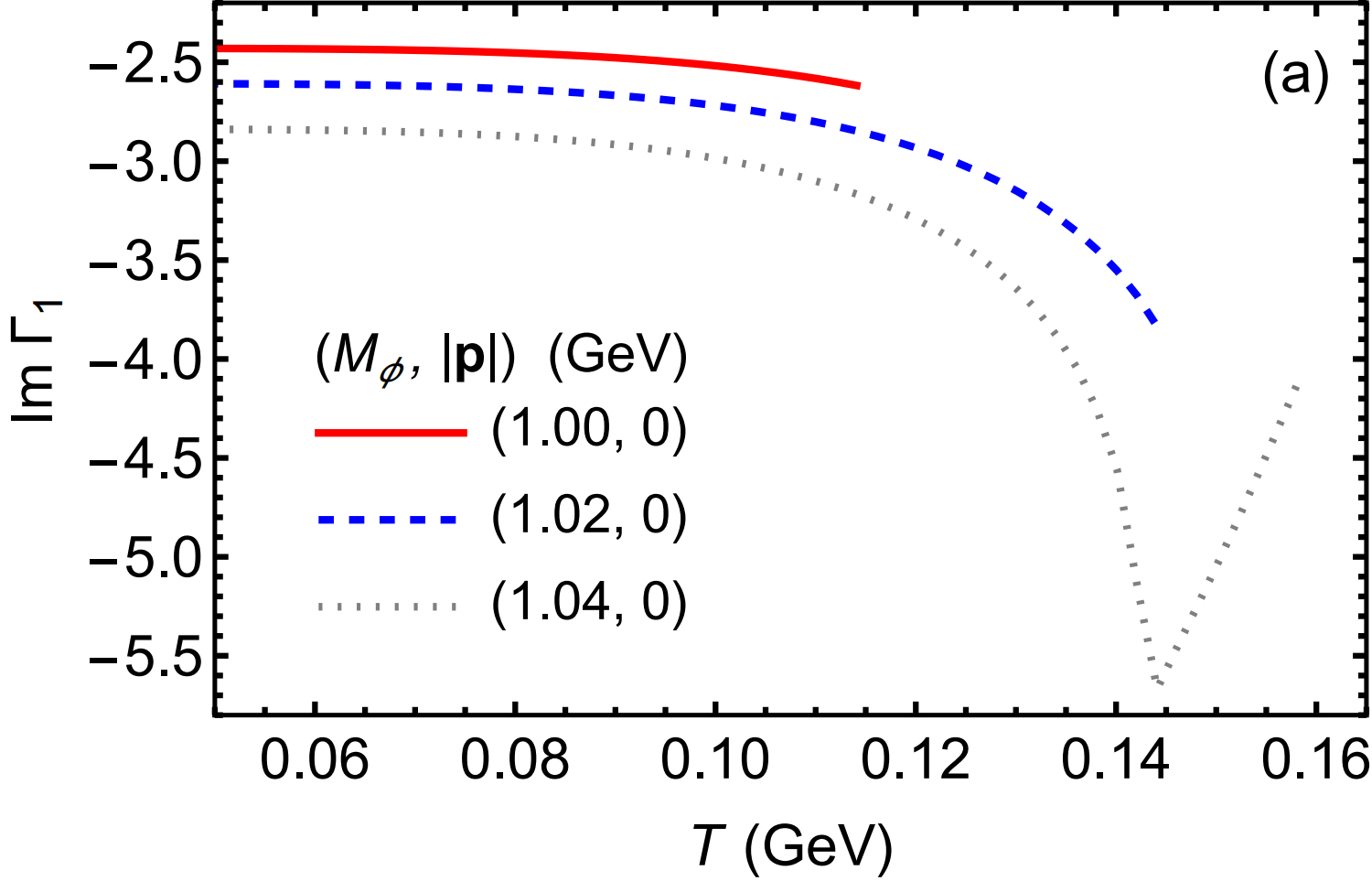}
    \caption{Real (right panel) and imaginary part (left panel) of $\Gamma_1(p_+,p_-)$ in Eq. (\ref{Gamma-decomposition}) as a function of temperature. Momenta $p_\pm^\mu$ are taken according to Eq. (\ref{eq:pp-pm-assumption}) and $M_\phi$ is set to 1 GeV (red solid line), 1.02 GeV (blue dashed line), and 1.04 GeV (gray dotted line).}
    \label{fig:gamma_1}
\end{figure}

Before we numerically calculate the spectral function, we now implement some simplifications for the contribution from kaon loop $\Pi_{\mu\nu}^\text{K-loop}$ in Eq. (\ref{self-energy}). An explicit calculation for $\Pi_{\mu\nu}^\text{K-loop}$ incorporates three closed loops: two from effective three-meson vertices and one from the kaon loop, making the calculation extremely complicated.  To simplify our calculation, we apply the dipole approximation (\ref{eq:dipole-approximation}) for the kaon propagator and note that the propagator is peaked around the mass-shell. Then what matters for our calculation is the case that both of the kaons' momenta in $\Pi_{\mu\nu}^\text{K-loop}$ are nearly on-shell. Focusing on the effective vertex in Eq. (\ref{eq:triangle-vertex}), we restrict ourselves to $p_+^2=p_-^2\approx M_K^2$ and $M_\phi=\sqrt{(p_++p_-)^2}\approx 1.02$ GeV. Here $M_K$ is a function of temperature and kaon's momentum as shown in Fig. \ref{fig:kaon-mass-coupling}. Then the kaon's momenta can be parameterized as follows,
\begin{align}\label{eq:pp-pm-assumption}
p_+^\mu+p_-^\mu&=\left(\sqrt{M_\phi^2+{\bf p}^2},\,0,\,0,\,|{\bf p}|\right)\,, \nonumber\\
p_+^\mu&= \left(E_+,\,|{\bf k}|\sin\theta,\,0,\,|{\bf p}|/2+|{\bf k}|\cos\theta\right)\,,
\end{align}
where $M_\phi$ denotes the $\phi$ meson's invariant mass and $\theta$ denotes the angle between $\phi$'s and kaon's three-momenta. The parameter $|{\bf k}|$ is determined by mass-shell conditions $p_+^2=M_K^2(p_+)$ and $p_-^2=M_K^2(p_-)$. We then show in Fig. \ref{fig:gamma_1} (a)
the real and imaginary parts of $\Gamma_1$ in Eq. (\ref{Gamma-decomposition}) as functions of the temperature. The coefficient $\Gamma_4=0$ for all the considered cases, while $\Gamma_2$ and $\Gamma_3$ are at least two orders of magnitude smaller than $\Gamma_1$,  allowing us to approximate the effective vertex as 
\begin{equation}\label{eq:approximation-gamma}
\tilde\Gamma^\mu(p_+,p_-)\approx (p_+^\mu-p_-^\mu)\Gamma_1(p_+,p_-)\equiv(p_+^\mu-p_-^\mu)\Gamma_\text{on}(M_\phi,|{\bf p}|)\,.
\end{equation}
We also checked that $\text{Im}\,\Gamma_1$ is nearly independent to $|{\bf p}|$, therefore in Fig. \ref{fig:gamma_1} (a) we only show results for $|{\bf p}|=0$. Since the kaon's mass is an increasing function of the temperature, we may have $\sqrt{M_\phi^2+{\bf p}^2}<2M_K$ if the temperature is high enough, indicating that the $\phi$ meson cannot be generated as a kaon resonance. As a consequence, the kaon loop does not contribute to the imaginary part of $\phi$ meson's self energy and thus the corresponding effective vertex is less important. This is why we did not show $\text{Im}\,\Gamma_1$ for $M_\phi=1$ GeV with $T>0.115$, $M_\phi=1.02$ GeV with $T>0.145$ GeV, and $M_\phi=1.04$ GeV with $T>0.161$ GeV in Fig. \ref{fig:gamma_1} (a). On the other hand, the real part of $\tilde{\Gamma}$ is nonzero only if $M_\phi> 2M_s$. Figure. \ref{fig:gamma_1} (b) shows $\text{Re}\,\Gamma_1$ as a function of temperature for several sets of $(M_\phi,|{\bf p}|)$. Specially, for the case of $M_\phi=1.04$ GeV and $|{\bf p}|=0$ GeV, we have a sudden change at $T=0.161$ GeV. This is because above this temperature, the $\phi$ meson production from combination of kaons is forbidden and thus we set $\text{Re}\Gamma_1=0$ by hand. 

Substituting the approximation (\ref{eq:approximation-gamma}) into Eq. (\ref{self-energy}), we derive $\Pi^\text{K-loop}_{\mu\nu}$ as follows
\begin{equation}
\Pi^\text{K-loop}_{\mu\nu}(p)=-8i \left|\Gamma_\text{on}(M_\phi,|{\bf p}|)\right|^2\int\frac{d^4 q}{(2\pi)^4}\frac{(2q_\mu-p_\mu)(2q_\nu-p_\nu)}{[q^2-M_K^2({\bf q})][(q-p)^2-M_K^2({\bf q-p})]}\,,
\end{equation}
where $M_\phi=\sqrt{p^2}$ is the invariant mass. Since $M_K$ is a slowly varying function of kaon's momentum, we further assume $M_K({\bf q})\approx M_K({\bf q-p})\approx M_K ({\bf p}/2)$ at finite temperature. These assumptions will greatly simplify the calculation for $\Pi^\text{K-loop}_{\mu\nu}$. They may not be significant enough at high temperature or high momentum region. However, in this paper, we restrict ourselves at $T\leq0.15$ GeV and $\phi$ meson's mass around its vacuum value, where the validity of the assumptions has been checked by our numerical calulations.  

\section{Numerical results for $K^+K^-$ pair production rate} \label{sec:numerical-results}

\subsection{Invariant mass spectrum in vacuum}

\begin{figure}[t]
    \centering
    \includegraphics[width=0.45\linewidth]{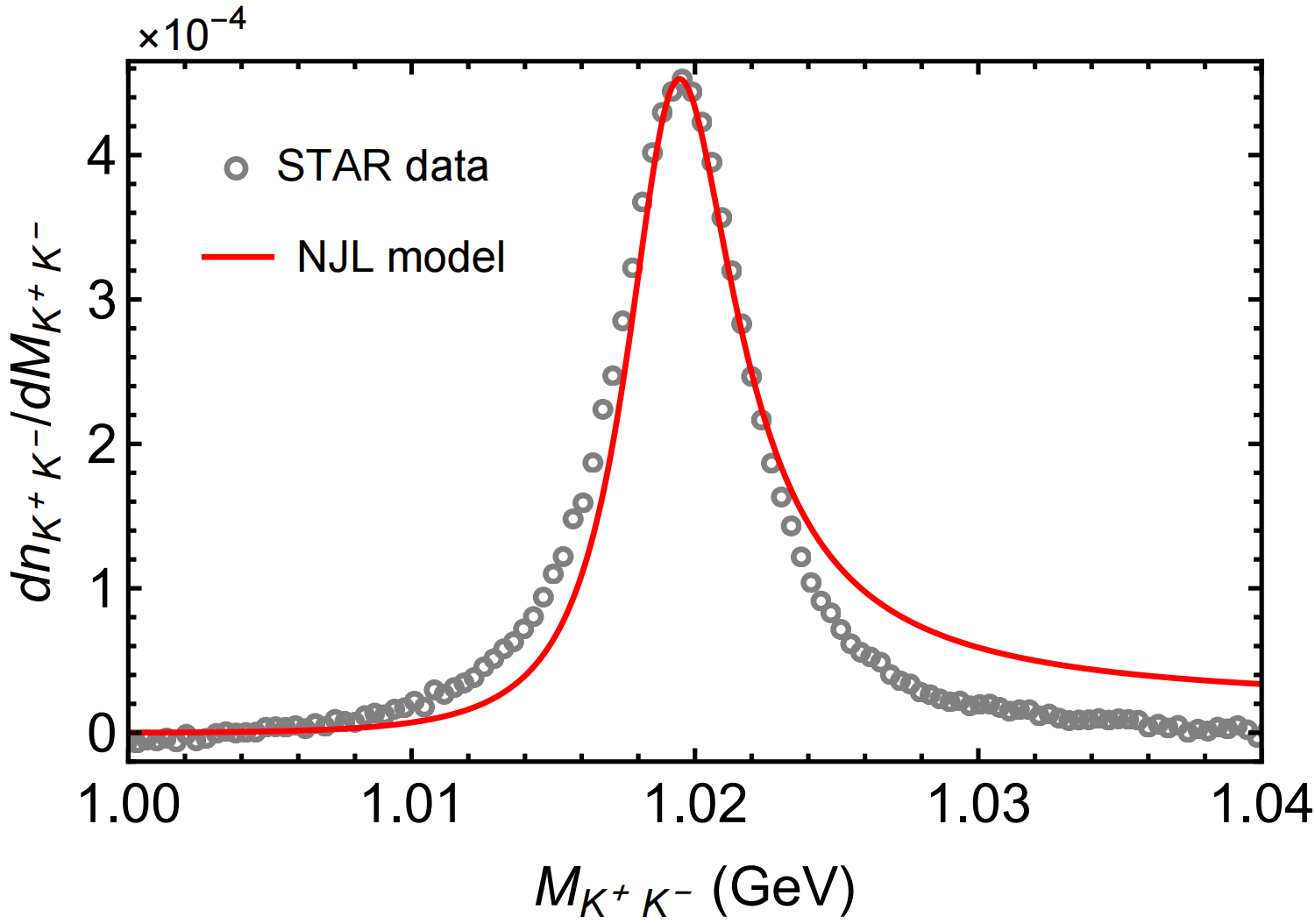}
    \caption{The invariant mass spectrum of the $K^+K^-$ pair calculated by NJL model in vacuum (red solid line), compared to the $K^+K^-$ pair yields measured by the STAR collaboration (gray circles) \cite{STAR:2022fan}. The experimental data is measured for Au+Au collisions at $\sqrt{s_\text{NN}}=27$ GeV at $20\%-60\%$ centrality, and summed over rapidity $|y|<1$ and transverse momentum $1.2\,\text{GeV}<p_T<1.8\,\text{GeV}$.}
    \label{fig:invariant-mass-vac}
\end{figure}

Within the framework of the NJL model, we now focus on the $K^+K^-$ pair in the vacuum. Since the vacuum is invariant under Lorentz transformation, the longitudinally and transversely polarized modes are degenerate and only depend on the invariant mass of the $K^+K^-$ pair. The differential production rate of $K^+K^-$ pair in (\ref{eq:KK-production-rate-new}) then reduce to the following invariant mass spectrum,
\begin{equation}
\frac{dn_{K^+K^-}}{dM_\phi}\propto \frac{1}{M_\phi}(M_\phi^2-4M_{K,\text{vac}}^2)^{3/2} \left|\Gamma_\text{on}^\text{vac}(M_\phi)\right|^2 \rho(p)\,,
\end{equation}
where we have used the approximation (\ref{eq:approximation-gamma}) for the meson's effective vertex. Here $M_\phi$ denotes the mass of the intermediate $\phi$ meson, or equivalently saying, the invariant mass of the final state $K^+K^-$ pair. The spectral function $\rho(p)$ is given by Eq. (\ref{eq:spectral-function-new}) with the self-energy determined by Eqs. (\ref{eq:self-energy-tot})-(\ref{eq:self-energy-tad}). Numerical results within the NJL model framework are shown in Fig. \ref{fig:invariant-mass-vac}, where we compare our results with the invariant mass spectrum observed in experiments \cite{STAR:2022fan}. Here we rescaled our result by a constant factor such that the peak value coincides with the experimental data. From Fig. \ref{fig:invariant-mass-vac} we observe that our calculation can nicely reproduce the spectrum of $K^+K^-$ pair. We observe that our NJL model calculation underestimates (overestimates) the invariant mass spectrum at energies below (above) the peak position. Consequently, the shape of our results deviates from the Breit-Wigner distribution. Such a deviation arises because the imaginary part of the self-energy is not a constant but a function of the invariant mass, as illustrated in Fig. 9 (a), whereas the width parameter in the Breit-Wigner distribution is assumed to be independent to the invariant mass.

\subsection{Invariant mass spectrum at finite temperature}

\begin{figure}[b]
    \centering
    \includegraphics[width=0.45\linewidth]{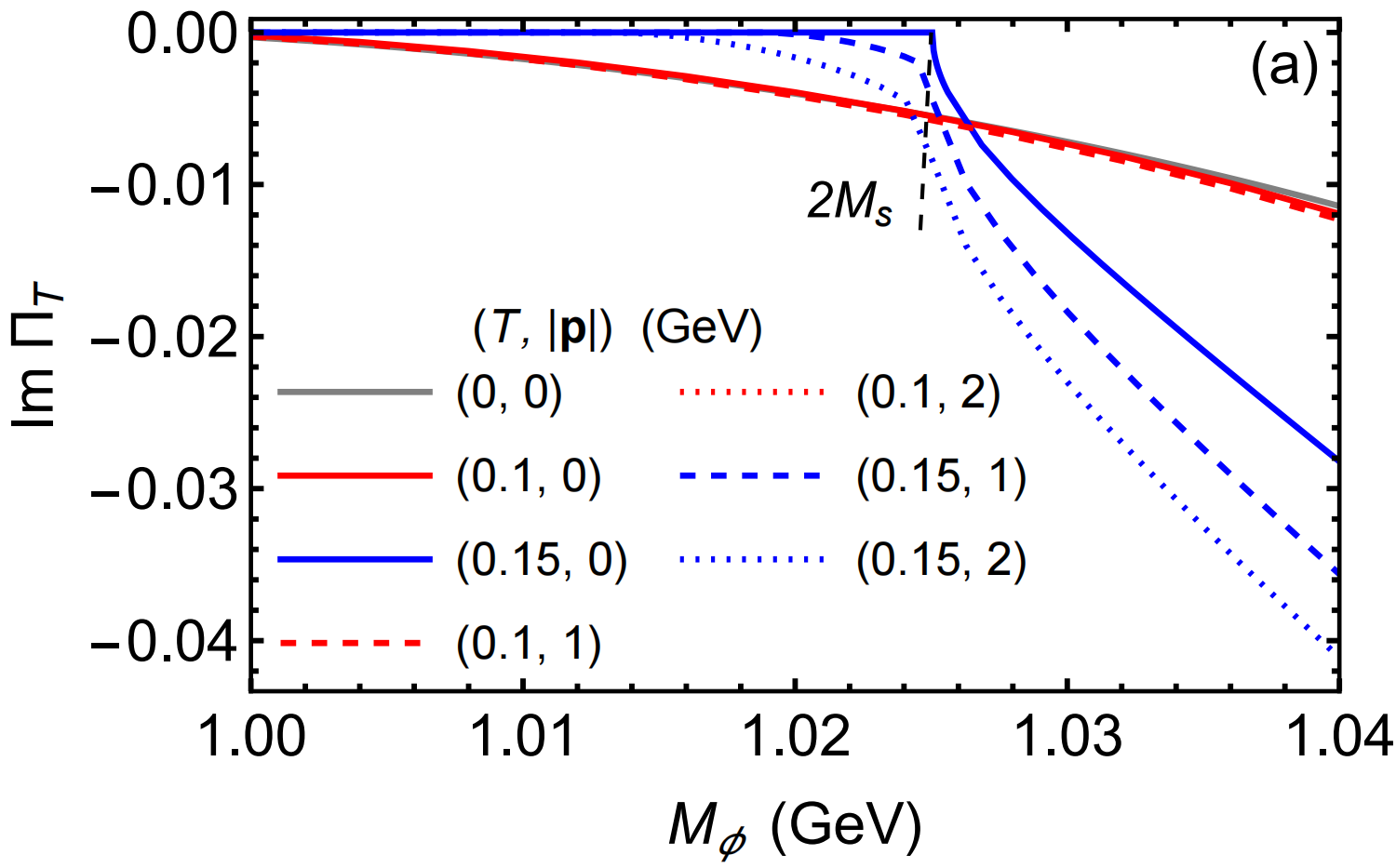}
    \includegraphics[width=0.413\linewidth]{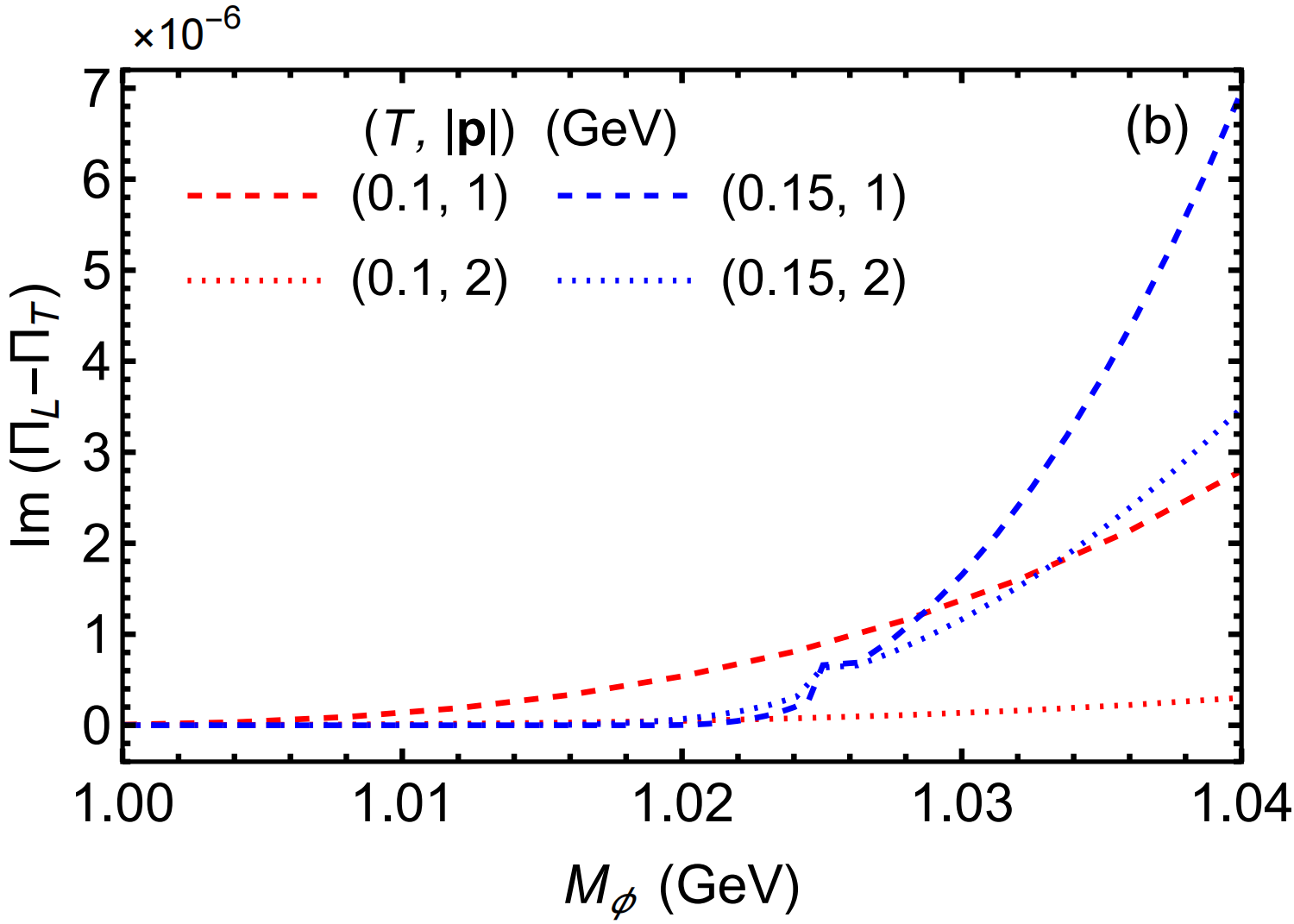}
    
    \caption{Left panel: the imaginary part of the $\phi$ meson's self-energy for transversely polarized mode as a function of meson's invariant mass. Right panel: $\text{Im}\,(\Pi_L^\text{tot}-\Pi_T^\text{tot})$ as a function of meson's invariant mass. We focus on the $\phi$ meson with momentum $|{\bf p}|=0$ (solid lines), 1 GeV (dashed lines), and 2 GeV (dotted lines) at temperatures $T=0$ (the gray line), $0.1$ GeV (red lines), and $T=0.15$ GeV (blue lines). In panel (a), the position of $2M_s$ at $T=0.15$ GeV is indicated by the dashed vertical line. }
    \label{fig:spectral-functions-T}
\end{figure}

At finite temperature, unlike in the vacuum, the spectral functions $\rho_L$ and $\rho_T$ are different if the $\phi$ meson (or equivalently saying, the $K^+K^-$ pair) has a nonzero velocity relative to the thermal background. Then the production rate of $K^+K^-$ pair in (\ref{eq:KK-production-rate-new}) is converted to 
\begin{equation}\label{eq:KK-production-rate-fT-df}
\frac{dn_{K^+K^-}}{dM_\phi d^{3}{\bf p}\,d\cos\theta^{*}}\propto f_T(p) + \delta f(p) \cos^2\theta^\ast\,,
\end{equation}
where we have introduced two auxiliary functions $f_T(p)$ and $\delta f(p)$. They are defined as follows,
\begin{equation}
\left(\begin{array}{c}
f_T(p)\\
\delta f(p)\\
\end{array}\right)\equiv\frac1\omega n_{B}(\omega) (M_\phi^2-M_{K,\text{vac}}^2)^{3/2}\left|\Gamma_\text{on}^\text{vac}(M_\phi)\right|^2
\left(\begin{array}{c}
\rho_{T}(p)\\
\rho_{L}(p)-\rho_{T}(p)\\
\end{array}\right)\,,
\end{equation}
where $\omega=\sqrt{M_\phi^2+{\bf p}^2}$ denotes the total energy of the $K^+K^-$ pair in the rest frame of the thermal background. According to Eq. (\ref{eq:spectral-function-new}), the spectral functions $\rho_{L/T}$ are affected by the temperature through the imaginary parts of self-energies, $\text{Im}\,\Pi^\text{tot}_{L/T}$. At finite temperature, these functions depend on the invariant mass $M_\phi$ and the momentum $|{\bf p}|$. We present $\text{Im}\,\Pi^\text{tot}_T$ and $\text{Im}\,(\Pi_L^\text{tot}-\Pi_T^\text{tot})$ as functions of $M_\phi$ in Fig. \ref{fig:spectral-functions-T} (a) and (b), respectively. We note that the transversely polarized mode degenerates with the longitudinally polarized mode when $|{\bf p}|=0$, leading to $\text{Im}\,(\Pi_L^\text{tot}-\Pi_T^\text{tot})=0$. However, $\text{Im}\,\Pi^\text{tot}_T$ and $\text{Im}\,\Pi^\text{tot}_L$ have nonvanishing values even if $|{\bf p}|=0$, as indicated by solid lines in Fig. \ref{fig:spectral-functions-T} (a). At temperature $T=0.1$ GeV, shown by the red solid, dashed, and dotted lines in Fig. \ref{fig:spectral-functions-T} (a), corresponding to $|{\bf p}|=0$, 1 GeV, and 2 GeV, respectively,  $\text{Im}\,\Pi^\text{tot}_T$ as a function of $M_\phi$ is nearly identical with the one at zero temperature (the gray solid line). This is because at low temperatures, e.g., $T\lesssim0.1$ GeV, the quark's and kaon's masses are almost independent to the temperature, as shown by Figs. \ref{fig:Quark-dynamic-mass} and \ref{fig:kaon-mass-coupling}. In this temperature region, only the kaon loop contributes to $\text{Im}\,\Pi^\text{tot}_T$, indicating that the intermediate $\phi$ meson is purely generated as a kaon-pair resonance. At a higher temperature, such as $T=0.15$ GeV, the quark loop starts to contribute to $\text{Im}\,\Pi^\text{tot}_T$ when $M_\phi>2M_s(T)$. As a consequence, the blue line in Fig. \ref{fig:Quark-dynamic-mass} (a) undergoes a sudden change at $2M_s(T)$. We also observe that $\text{Im}\,\Pi^\text{tot}_T$ is negative definite, thus the spectral function in Eq. (\ref{eq:spectral-function}) is positive definite. On the other hand, from Fig. \ref{fig:spectral-functions-T} (b) we find that the difference between $\text{Im}\,\Pi^\text{tot}_L$ and $\text{Im}\,\Pi^\text{tot}_T$ is 4 orders of magnitude smaller than $\text{Im}\,\Pi^\text{tot}_T$, which is also much smaller than predictions from other models \cite{Kim:2019ybi,Li:2022vmb,Dong:2023cng}.  This means that the transversely and longitudinally polarized modes have nearly the same property in the framework of the NJL model.

\begin{figure}[t]
    \centering
    \includegraphics[width=0.45\linewidth]{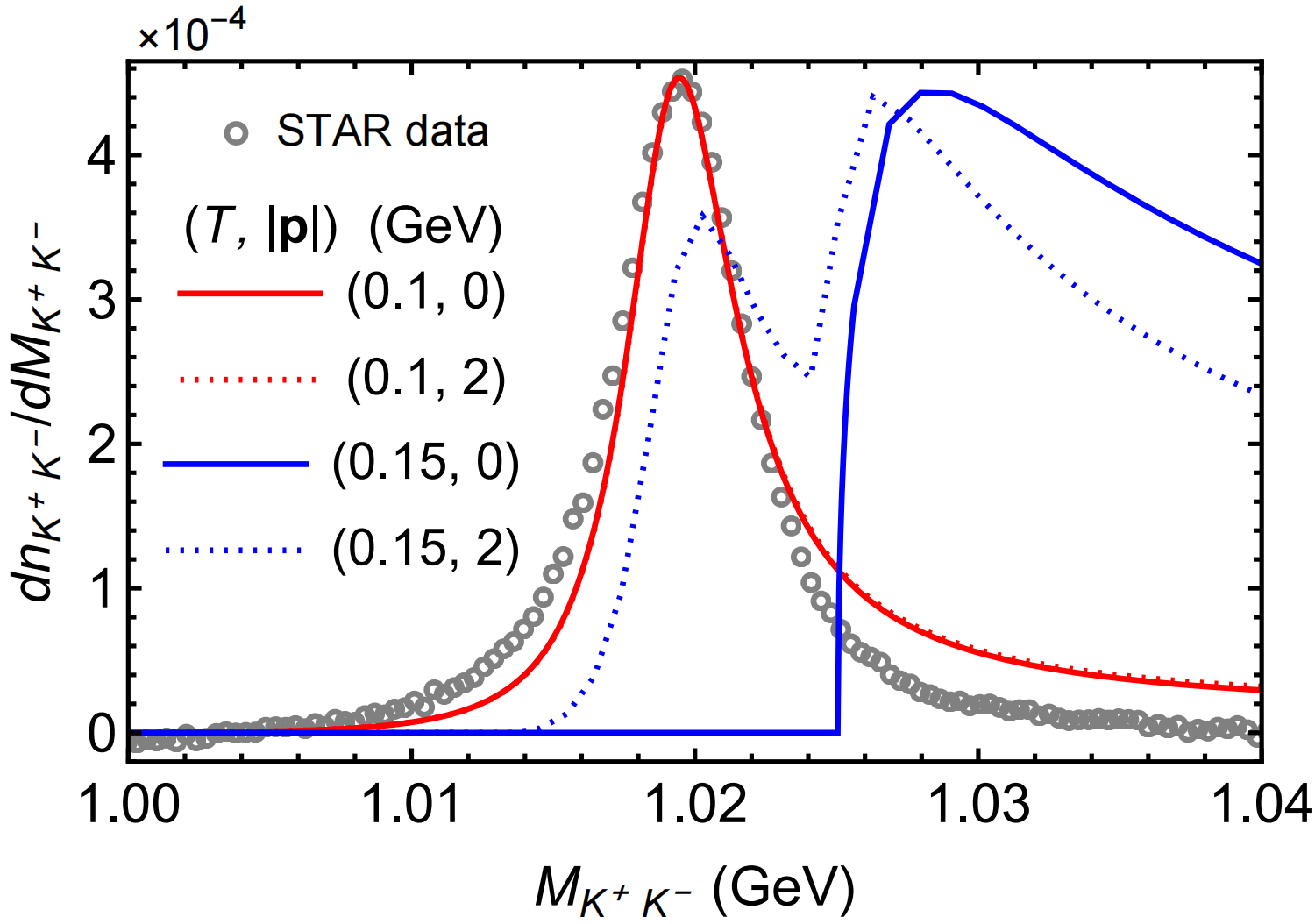}
    
    \caption{Invariant mass spectrum for $K^+K^-$ pair at finite temperature. Labels are taken in the same way as Fig. \ref{fig:spectral-functions-T}.}
    \label{fig:invariant-mass-T}
\end{figure}

Using the self energy of $\phi$ meson, we show the invariant mass spectrum of $K^+K^-$ pair at different sets of temperature $T$ and momentum $|{\bf p}|$ in Fig. \ref{fig:invariant-mass-T}. At $T=0.1$ GeV, the spectral function, shown by the red solid and dotted lines, is nearly independent to $|{\bf p}|$ and is almost the same as the spectrum in vacuum in Fig. \ref{fig:invariant-mass-vac}. This behavior is again attributed to the weak $T$- and $|{\bf p}|$-dependence of the quark's and kaon's masses in the region of $T\lesssim0.1$ GeV. At temperature $T=0.15$ GeV, the location of the peak arises a significant positive shift. When the momentum $|{\bf p}|=0$, only the imaginary part of the quark loop contribution has a nonvanishing value in the considered invariant mass region and the spectral function shows a much broader width compared to the vacuum case. 
On the other hand, when $|{\bf p}|=2$ GeV, both the kaon loop and the quark loop contribute to $\text{Im}\,\Pi^\text{tot}_{T/L}$, leading to a double-peaked structure, shown by the blue dotted line.  

\section{Numerical results for $\phi$ meson's spin alignment}\label{sec:numerical-spin}

\begin{figure}[t]
    \centering
    \includegraphics[width=0.45\linewidth]{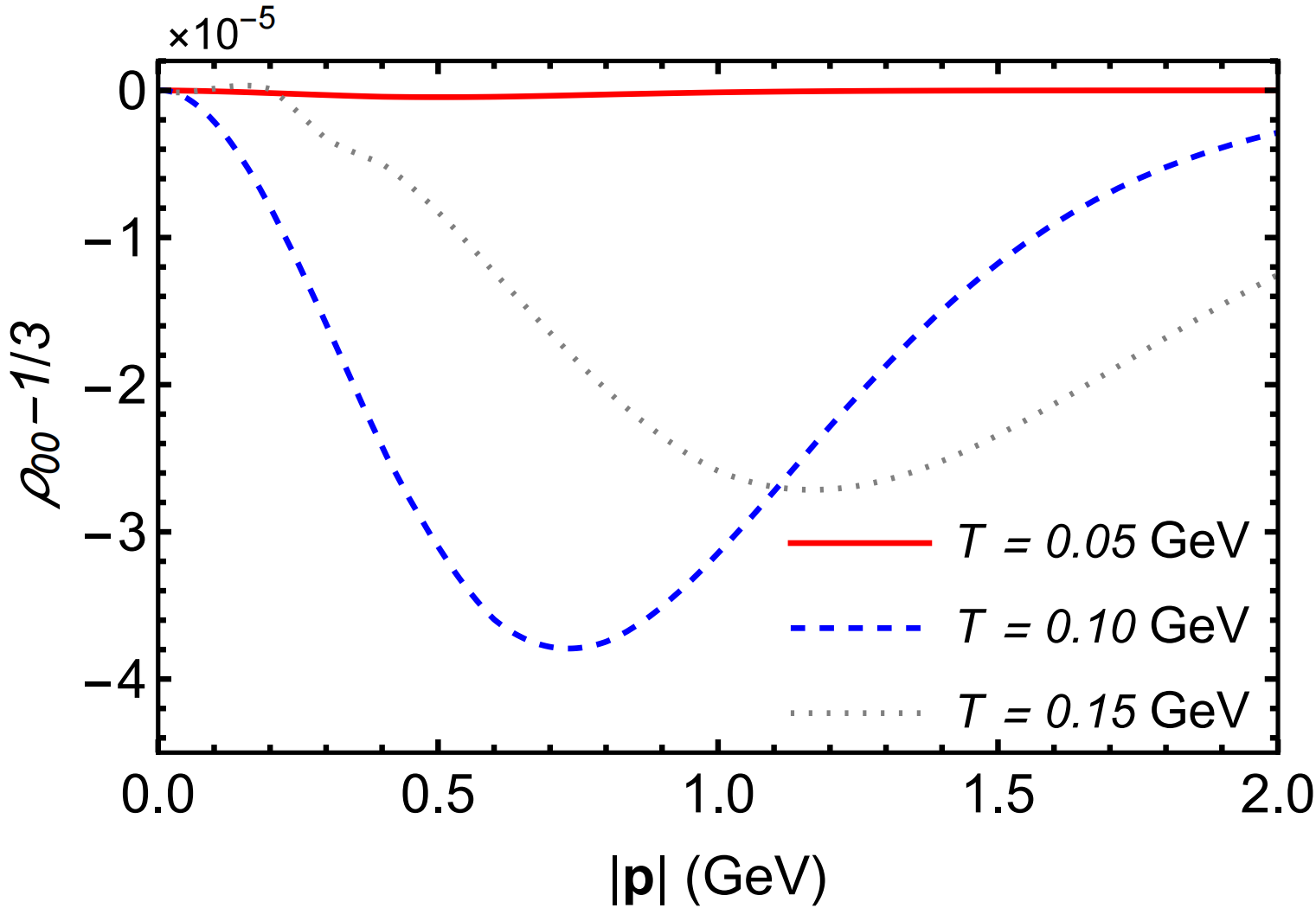}
    
    \caption{The spin alignment of the $\phi$ meson as a function of the momenta $|{\bf p}|$ relative to the thermal background. The $\phi$ meson is produced at $T=0.05$ GeV (red solid line), 0.1 GeV (blue dashed line), or 0.15 GeV (gray dotted line).}
    \label{fig:spin-alignment}
\end{figure}

In this section we focus on the $\phi$ meson's spin alignment. In practice, what is usually used is the spin alignment averaged over a given region of invariant mass, transverse momentum, and rapidity. By fixing the momentum $|{\bf p}|$ while averaging over the invariant mass region $M_\text{min}<M_\phi<M_\text{max}$, we introduce the average spin alignment $\overline{\rho}_{00}({\bf p})$ that satisfies
\begin{equation}
\int_{M_\text{min}}^{M_\text{max}}dM_\phi\frac{dn_{K^+K^-}}{dM_\phi d^3{\bf p}\,d\cos\theta^\ast}\propto\left[1-\overline{\rho}_{00}({\bf p})\right]+\left[3\overline{\rho}_{00}({\bf p})-1\right]\cos^2\theta^\ast\,.
\end{equation}
Compared to Eq. (\ref{eq:KK-production-rate-fT-df}), we obtain that 
\begin{equation}
\overline{\rho}_{00}({\bf p})-\frac13=\frac{2\int_{M_\text{min}}^{M_\text{max}} dM_\phi \delta f(p)}{3\int_{M_\text{min}}^{M_\text{max}} dM_\phi \left[3f_T(p)+\delta f(p)\right]}\,.
\end{equation}
If $\rho_L=\rho_T$, we have $\delta f=0$ and thus $\overline{\rho}_{00}({\bf p})=1/3$, as expected. In the framework of the NJL model, we take $M_\text{min}=1$ GeV and $M_\text{max}=1.04$ GeV and present in Fig. \ref{fig:spin-alignment} the average spin alignment as functions of $|{\bf p}|$ and temperature $T$. The spin alignment first decreases and then increases as $|{\bf p}|$ increases. At $T=0.1$ GeV, $\bar{\rho}_{00}$ reaches its minimum value at $|{\bf p}|=0.72$ GeV, while at $T=0.15$ GeV the minimum value corresponds to a larger $|{\bf p}|$. Such a behavior indicates that the motion of $\phi$ meson relative to the thermal background breaks the symmetry between longitudinally and transversely polarized states, but this symmetry will be restored at sufficiently large momentum. Deviations of $\overline{\rho}_{00}({\bf p})$ from $1/3$ are negative, indicating that the longitudinally polarized modes have smaller probabilities. However, the magnitude of the deviations is just $10^{-5}$ at $T=0.1$ or 0.15 GeV, and even smaller at $T=0.05$ GeV. This is because in this paper we only include the chiral condensate, which keeps the translation invariance. Taking into account tensor condensates could induce a larger spin alignment, in analogue to the strong force field proposed in Ref. \cite{Sheng:2019kmk,Sheng:2022wsy}, which will be addressed in future studies. 

\section{Summary}

The experimental observation of the $\phi$ meson's spin alignment in heavy-ion collisions \cite{STAR:2022fan} has attracted interests in studying the $\phi$ meson's in medium properties. In experiments, the $\phi$ meson is reconstructed via its strong decay $\phi\rightarrow K^++K^-$, while its spin alignment is determined by the polar angle distribution of the daughter $K^+$ meson's momentum in the $\phi$ meson's rest frame. In this paper, we focus on the process that a $\phi$ meson is formed in the QGP from a quark-antiquark pair (or a pair of kaons) and subsequently decays to $K^+K^-$ outside the QGP region. Analytical expression for the differential production rate of final state $K^+K^-$ pair is derived, which is related to the $\phi$ meson's self-energy and propagator. We then parameterize the production rate using the four-momentum of $K^+K^-$ pair, the azimuthal angle, and the polar angle with respect to a spin quantization direction. For a $\phi$ meson propagating in an isotropic thermal background, a natural choice for the spin quantization direction is the momentum direction, corresponding to the helicity frame. Then the $K^+K^-$ production rate, and consequently the spin alignment, are related to spectral functions of longitudinally and transversely polarized $\phi$ mesons.

Based on the $SU(3)$ NJL model, we present a detailed calculation for the $K^+K^-$ production through the $\phi$ meson's decay. We take into account the chiral condensate and regularize the model with a hard cutoff to avoid ultraviolet divergences in momentum integrals. The NJL model used in this paper keeps the isospin symmetry between $u$ and $d$ quarks, and therefore $K^\pm$, $K^0$, and $\overline{K}^0$ have the same mass, which is determined by the poles of the kaon propagator, with the kaon's self-energy attributed by a quark loop. On the other hand, the crucial point for evaluating the $K^+K^-$ production rate is the self-energy of the $\phi$ meson. With this self-energy, it is straightforward to obtain the propagator by solving the Dyson-Schwinger equation. In this work, we extend the calculation of self-energy to include contributions of kaon loops, which are next-to-leading order corrections in the $1/N_c$ expansion. Phenomenologically, this incorporates the $\phi$ meson as a $K^+K^-$ resonance, which is also a bound state of $s\bar{s}$. We then numerically calculated the production rate of the final-state $K^+K^-$ pair. 
At zero temperature, the invariant mass spectrum calculated in our model shows a very good agreement with the experimental data~\cite{STAR:2022fan}, suggesting that the kaon loop plays the crucial role for describing the vacuum properties of the $\phi$ meson within the framework of the NJL model. We also find that the spectrum is nearly independent to the temperature when $T\lesssim 0.1$ GeV. At a higher temperature, e.g., $T=0.15$ GeV, the quark coalescence contributes to the invariant mass spectrum of $K^+K^-$ pair, leading to a much broader width which is inconsistent with experiment. 

The inclusion of kaon loop contribution to the $\phi$ meson's self-energy is the key point for reproducing the invariant mass spectrum of $K^+K^-$ pairs. If we only consider the quark loop contribution, as did in many previous papers \cite{Klevansky:1992qe,Rehberg:1995kh,Sheng:2022ssp}, the $\phi$ meson in vacuum will be a stable $s\bar{s}$ bound state, whose invariant spectral function is a $\delta$-function. The kaon loop, on the other hand, opens the physical decay channels $\phi\rightarrow K^+ +K^-$ and $\phi\rightarrow K^0_L+K^0_S$ and thus the $\phi$ meson is no longer stable, as reflected by a finite width in the spectrum. 

The spin alignment of the $\phi$ meson, quantified by $\rho_{00}-1/3$, is found to be of the order of $10^{-5}$ in our thermal background calculation, significantly smaller than the experimental measurements \cite{STAR:2022fan}. This discrepancy suggests that additional mechanisms, such as tensor condensates, in analogue to the strong force field proposed in   \cite{Sheng:2019kmk,Sheng:2022wsy,Sheng:2023urn}, may play a dominant role in the spin alignment. Other external fields, e.g., vorticity or magnetic fields, could also have nontrivial contributions to the spin alignment. Our current framework, which includes only the Lorentz-invariant chiral condensate, does not capture these effects. Future studies should explore the influence of such non-perturbative backgrounds to better understand the observed spin alignment.

In summary, our results highlight the importance of kaon loop contributions in describing the $\phi$ meson’s spectral properties and underscore the necessity for further theoretical developments to explain its spin alignment. These findings pave the way for future investigations into the interplay between QGP dynamics and hadronic observables.

\section*{Acknowledgment}

The authors thank Marcus Bleicher, Hai-Cang Ren, Dirk H. Rischke, and Shu-Yun Yang for helpful discussions. D.H. and X.N.Z. are supported by the National Key Research and Development Program of China under Contract No. 2022YFA1604900, and by the National Natural Science Foundation of China (NSFC) under Grant No.12435009 and No. 12275104. X.N.Z. is also supported by China Scholarship Council under Grant No. 202406770002. X.L.S. acknowledges the financial support by the Italian Ministry of University and Research, project PRIN2022 "Advanced probes of the Quark Gluon Plasma", funded by Next Generation EU, Mission 4 Component 1.
 
\bibliographystyle{apsrev4-1}
\bibliography{bibfile}

\end{document}